\documentclass[showpacs,aps,prl,twocolumn,superscriptaddress]{revtex4-1}

\usepackage{graphicx}
\usepackage{hyperref}
\usepackage{epstopdf}
\usepackage{dcolumn}% Align table columns on decimal point
\usepackage{bm}% bold math
\usepackage{amsmath}
\usepackage{amssymb}
\usepackage[dvipsnames]{xcolor}
\usepackage{color} 
\usepackage{ulem}
\usepackage{xspace}
\usepackage{ulem}

\usepackage{float}

\renewcommand{\vec}[1]{\boldsymbol{#1}}

\newcommand{\ve}[1]{\boldsymbol{#1}}

\setcitestyle{square,numbers}
\usepackage{here}

\begin{document}

\title{Quantum Monte Carlo Simulation of Generalized Kitaev Models}

\author{Toshihiro Sato}
\affiliation{\mbox{Institut f\"ur Theoretische Physik und Astrophysik, Universit\"at W\"urzburg, 97074 W\"urzburg, Germany}}
\author{Fakher F. Assaad}
\affiliation{\mbox{Institut f\"ur Theoretische Physik und Astrophysik, Universit\"at W\"urzburg, 97074 W\"urzburg, Germany}}
\affiliation{\mbox{W\"urzburg-Dresden Cluster of Excellence ct.qmat, Am Hubland, 97074 W\"urzburg, Germany}}

\begin{abstract}
Frustrated spin systems generically suffer from the negative sign problem inherent to Monte Carlo methods.  Since the  severity of this problem is formulation dependent, optimization   strategies can be put forward.    We introduce a phase pinning approach  in the realm of the auxiliary field  quantum Monte Carlo algorithm.  If  we can find an  anti-unitary operator that commutes  with 
the  one body Hamiltonian coupled to the  auxiliary field,  then the phase of the action is pinned to  $0$ and $ \pi$.  For generalized Kitaev models,  we  can successfully apply  this strategy  and observe a remarkable improvement of  the  average sign.  We use this method to study thermodynamical  and  dynamical  properties of the  Kitaev-Heisenberg model   down to temperatures  corresponding to  half of the exchange coupling constant.  Our dynamical data reveals   finite temperature properties of   ordered and  spin-liquid phases inherent to this model. 
\end{abstract}

\maketitle

{\it Introduction.}---   
Local moment formation and  spin-orbit entanglement  is at the origin of many fascinating states of matter  that are realized in various materials \cite{Kim_rev2013}.  The family of layered 
iridates and $\alpha$-RuCl$_3$  are  Mott insulators   where strong spin-orbit coupling   leads to bond selective  spin couplings  on an  underlying honeycomb lattice  \cite{Jackeli-KSL-01,Chaloupka-2013, Plumb2014}.  This class of materials is believed to be proximate to the Kitaev spin liquid characterized by emergent   Majorana fermions and $Z_2$ fluxes   \cite{Kitaev06}.   In particular, $\alpha$-RuCl$_3$   exhibits  zig-zag spin ordering, but  proximity to the Kitaev spin liquid suggests that high energy features of this material   are described by  Majorana fermions \cite{Do:2017aa,Banerjee1055}.   These exotic particles will hence only show up in thermaldynamical and dynamical  properties   in an intermediate temperature range  bounded by the ordering temperature and the  coherence  scale of the Majorana fermions. 

\begin{figure}[t]                              
\begin{center}    
\includegraphics[scale=0.45]{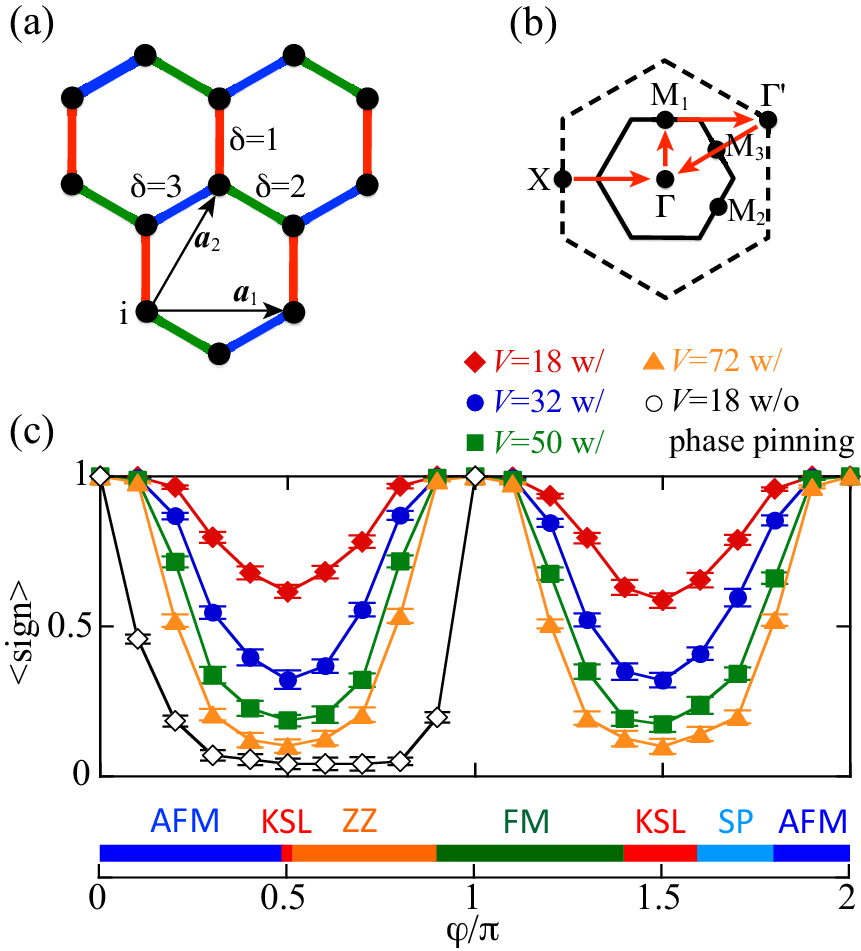} 
\caption[]{
(a)  Spin 1/2  degrees of freedom, $\hat{\ve{S}}_{\vec{i}} $,  on the honeycomb lattice are subject to  
 Heisenberg  $J \hat{\ve{S}}_{i}  \cdot \hat{\ve{S}}_{i + \delta}$ 
and Kitaev   $2K\hat{S}_{i}^{\delta}  \hat{S}_{i+\delta}^{\delta}$  exchange interactions.   Here  $\delta=1$(red), $2$(green), and $3$(blue) runs over the there bonds, and 
$\boldsymbol{a_1}$ and $\boldsymbol{a_2}$  correspond to the lattice vectors. 
 (b) First (solid) and second (dashed line) Brillouin zones.
(c) Average sign $\langle \text{sign} \rangle$ as a function of $V$ for  various angles $\varphi$.   The figure includes the  ground-state phase diagram with antiferromagnetic (AFM), Kitaev spin liquid (KSL), zig-zag (ZZ), ferromagnetic (FM), and stripy (SP)  phases, as proposed in Ref.~\cite{Chaloupka-2013}.     
Here  we have set the temperature to $T=1$ in units of  $A$. 
}  
\label{fig:Fig1}                                     
\end{center}
\end{figure}

The aim of this Letter is to  provide a quantum Monte Carlo (QMC) algorithm that allows one to study   a  generalized  Kitaev model  in a temperature range that overlaps with the aforementioned energy scales.  
For concreteness, we consider:
\begin{eqnarray}
\label{Eq:KHM}
\hat{H}   =  \sum_{i,j,\alpha,\beta}  \Gamma_{i,j}^{\alpha,\beta} \hat{S}_{i}^{\alpha}  \hat{S}_{j}^{\beta}  + \sum_{i,j}  J_{i,j}  \hat{\ve{S}}_{i}  \cdot \hat{\ve{S}}_{j}.
\end{eqnarray}
Here  $i,j$ run over  sites of the honeycomb lattice  and 
$\hat{S}_{i}^{\alpha} $ is a spin 1/2 degree of freedom.    
For  $i,j$   defining a nearest neighbor $\delta$-bond  (see Fig.~\ref{fig:Fig1}(a))   and $ \Gamma_{ \delta}^{\alpha,\beta } =  2K \delta_{\alpha,\beta} \delta_{\delta,\alpha} $ the first  term reduces to the Kitaev model~\cite{Kitaev06}.  Although redundant,  it is  convenient for the  simulations to include an $SU(2)$-symmetric Heisenberg term  with non-frustrating  exchange  couplings $J_{i,j}$.  

Hamiltonians of the form in Eq.~(\ref{Eq:KHM}) suffer from the negative sign problem such that no exact QMC simulations have been carried out to date. 
Numerical research for this class of Hamiltonians  has made use of  exact diagonalization~\cite{Chaloupka2010,Chaloupka-2013, Katukuri_2014,Yamaji-1,Winter:2017aa,Winter2018,Laurell:2020aa}, functional renormalization group~\cite{Reuther2011,Singh2012}, density-matrix renormalization group~\cite{Gohlke-2017}, and  the  thermal pure quantum state method~\cite{Yamaji-1,Laurell:2020aa}.
The  negative sign problem in the  QMC approach is formulation dependent and hence can, in principle, be  reduced so as to reach relevant  energy scales. In fact, this can be seen as an optimization problem over the space  of possible path integral formulations \cite{Wan20,Hangleiter20}.     Here we adopt  a  symmetry based strategy,  that pins the phase of the action to  0 and $\pi$.    We will show that this  strategy  greatly reduces the severity of the negative sign problem and that it opens a  window of temperatures  where the QMC  works efficiently and that is relevant to experiments. 

{\it Phase pinning approach.}---
The   auxiliary field QMC (AFQMC) algorithm~\cite{Blankenbecler81,White89,ALF_v1} is based on a Hubbard-Stratonovich decoupling of the  interaction term. After this step,  the  partition function can  generically be written as 
\begin{equation}
	   Z  = \int d \Phi(x,\tau) e^{- S (\Phi(x,\tau)  )} 
\end{equation} 
with  
\begin{equation}
	   S(\Phi)    =  S_0(\Phi) - \log   \text{Tr} \left[ {\cal T}   e^{-\int_{0}^{\beta}  d \tau   \sum_{x,y}\hat{c}^{\dagger}_x  h_{x,y}(\tau)   \hat{c}^{\phantom\dagger}_y } \right] .
\end{equation} 
Here, $\Phi$ corresponds to the Hubbard-Stratonovich field,  $\hat{c}^{\dagger}_{x}$  are  fermion operators, $x$  runs   over the  single particle states, $S_0$ is a  real \textit{bosonic} action and  $ h_{x,y}(\tau)  $ is a $\Phi$  and $\tau$ dependent matrix.
The trace over the fermion degrees of freedom is generically complex  such that  the phase,  $ \text{Im}  S   \in [0, 2 \pi ]$.   The Monte Carlo  importance sampling of the field $\Phi$  is then carried out  according to    weight $ 	|e^{-S(\Phi)} | $  and the average sign corresponds  to the reweighting factor  $\langle   \text{sign}  \rangle   =  \int d \Phi e^{- S (\Phi )}  /  \int d \Phi  | e^{- S (\Phi )} |   $.     Generically, the  average sign scales as $e^{- \Delta \beta V } $   with $V$ the volume of the system  and $ \Delta $ a formulation dependent constant. Since  the errors on the average sign have to be smaller than the mean value,   the computational  cost  required to resolve this quantity scales as $e^{2 \Delta \beta V } $.     Within the above framework,  the sign problem amounts to the fluctuations of the phase. 
Using symmetry considerations \cite{Wu04,Wei16,Li16}  one can show that  one can pin the phase to  $\text{Im} S = 0$ thus solving the sign problem. 
For instance, in Ref.~\cite{Li16}, it is shown that the negative sign problem is absent if one can find two anti-unitary  operators  that mutually anti-commute and that commute with 
$h(\tau)$.     This insight has  greatly  enhanced the class of sign free model Hamiltonians 
 \cite{SatoT17_1,Liu18,Ippoliti18,WangZ20,Schattner15,Pan20,Huffman14} that one can simulate with the AFQMC.    For many models, no sign free formulations are know.  The question then arises:  how should optimize the sign by minimizing $\Delta$?     We will follow the idea that   reducing the fluctuations of     $\text{Im}S$  will reduce the severity of the sign problem.    In  particular  if we  can design a formulation of the path integral   such that there  exits a  single anti-unitary operator that commutes with 
$h(\tau)$ then the phase is pinned to $ \text{Im}S = 0,\pi$.      A proof of this statement is given in the Supplemental Material. 
Note that for the doped Hubbard model where  formulations can be found with ${\rm{Im}}S = 0, \pi $, many interesting high temperature properties have been studied \cite{Huang18,Huang19}.  

The generalized Kitaev model of Eq.~(\ref{Eq:KHM})   falls into this category. The first step is  to adopt a fermion representation of the spin-1/2 degree of freedom:  $ \hat {\ve{S}} = \frac{1}{2} \hat{\ve{f}}^{\dagger}   \hat{ \ve{\sigma}}  \hat{\ve{f}}^{\phantom\dagger}$ where $\hat{\ve{f}}^{\dagger}  \equiv  (\hat {f}^{\dagger}_{\uparrow}, \hat f^{\dagger}_{\downarrow} ) $ is a two-component fermion with constraint $\hat{\ve{f}}^{\dagger} \hat{\ve{f}}^{\phantom\dagger}   = 1$.    We then consider the Hamiltonian
\begin{eqnarray}
\label{Eq:HQMC}
\hat{H}_{{\rm QMC}}    &= & \sum_{ i,j, \alpha, \beta}\frac{|\Gamma_{i,j}^{\alpha,\beta} |}{2}  \left( \hat{S}_{i}^{\alpha}  + \frac{\Gamma_{i,j}^{\alpha,\beta}} {|\Gamma_{i,j}^{\alpha,\beta} |}   \hat{S}_{j}^{\beta}  \right)^2   \nonumber \\
& & - \sum_{ i,j}  \frac{J_{i,j}}{8}  \left(  \left(  \hat{D}^{\dagger}_{i,j}  + \hat{D}^{\phantom \dagger}_{i,j}  \right)^2+ \left(i \hat{D}^{\dagger}_{i,j}   -i \hat{D}^{\phantom\dagger}_{i,j}  \right)^2  \right) \nonumber \\
& &+U  \sum_{i} \left( \hat{\ve{f}}^{\dagger}_i \hat{\ve{f}}^{\phantom\dagger}_i  - 1 \right)^2,
\end{eqnarray}
where $ \hat{D}^{\dagger}_{i,j} = \hat{\ve{f}}^{\dagger}_i \hat{\ve{f}}^{\phantom\dagger}_{j} $.     
 It is important to note that 	$\left[  \left(  \hat{\ve{f}}^{\dagger}_{i}  \hat{\ve{f}}^{\phantom\dagger}_{i}  -1  \right)^2 ,  \hat H_{\rm{QMC}}   \right]  = 0 $ such that the  $\hat{\ve{f}}$-fermion parity  $ (-1)^{\hat{\ve{f}}^{\dagger}_{i} \hat{\ve{f}}^{\phantom\dagger}_{i} }  $  is a local conserved quantity  and that the constraint is very efficiently imposed.   In the odd parity sector favored by the  repulsive  Hubbard interaction, $ \left. \hat{H}_{\rm{QMC}} \right|_{(-1)^{ \hat{\ve{f}}^{\dagger}_{i}  \hat{\ve{f}}^{\phantom\dagger}_{i} }   = -1 }= \hat{H} + C $ where $C$ is a constant.
The  perfect squares can be   decomposed  with a standard Hubbard-Stratonovich 
decomposition.    Since the  $J_{i,j}$ couplings are non-frustrating,  we can find a set of Ising spins, $s_i = \pm 1 $,  such that  
 $J_{i,j} s_i s_j < 0$   for  all  bonds with  $| J_{i,j}| \neq 0 $.    One will  then show that for  each Hubbard-Stratonovich 
 configuration,   the single body propagator commutes  with the anti-unitary transformation:   $ \hat{T} \alpha \hat{f}^{\dagger}_{\ve{i},\sigma} \hat{T}^{-1} =  \overline{\alpha}  s_i  \hat{f}^{\phantom\dagger}_{\ve{i},\sigma} $.   The details of the calculation is presented in the Supplemental Material. 
 Thereby,  in this formulation, the phase is pinned to  $\text{Im} S = 0,\pi$.  

 {\it Case study.}---
 For concreteness,  we consider on each $\delta$-bond, $ \Gamma_{\delta}^{\alpha,\beta } =  2K \delta_{\alpha,\beta} \delta_{\delta,\alpha} $   and   $J_{\delta} = J $ in Eq.~(\ref{Eq:KHM}) (see Fig.~\ref{fig:Fig1}(a))   to obtain the Kitaev-Heisenberg model: 
\begin{eqnarray}
\label{Eq:KHM-2}
\hat{H}   =  2K \sum_{i  \in A ,\delta} \hat{S}_{i}^{\delta}  \hat{S}_{i+\delta}^{\delta}  + J \sum_{i \in A  , \vec{\delta} } \hat{\ve{S}}_{i}  \cdot \hat{\ve{S}}_{i + \delta}.
\end{eqnarray}
Here $i$ runs over the A sublattice and $i+\delta$ with $\delta=(1, 2, 3)$ over the  nearest neighbors.
The first term reduces to the Kitaev model~\cite{Kitaev06}.  
At  $K=0$  the  SU(2) spin symmetry of the  Heisenberg model allows for sign free AFQMC simulations (see Supplemental Material).  At any finite values of $K$  this symmetry is reduced to a $Z_2$ one  in which 
$  \ve{\ve{S}}_{\ve{i}}  \rightarrow -\ve{\ve{S}}_{\ve{i}}$ and no sign free formulation is known.   
We used the ALF (Algorithms for Lattice Fermions) implementation \cite{ALF_v1} of the well-established finite-temperature AFQMC method~\cite{Blankenbecler81,Assaad08_rev}  and   adopt the 
parametrization  $K=A{\rm{sin}}(\varphi)$,  $J=A{\rm{cos}}(\varphi)$,  with  $A=\sqrt{K^2+J^2}$. 
Henceforth, we use $A=1$ as the energy unit.
 Figure~\ref{fig:Fig1}(c) plots the average sign as a function of the angle $\varphi$   with and  without the phase pinning strategy.   One observes a remarkable  improvement of the average sign  when the phase is pinned   to $0 ,\pi$.  
A  crucial question is if we can reach experimental relevant energy scales for Kitaev materials. 
Typical energy scales such as the charge gap $\Delta_{\rm c}$~\cite{Winter_2017} and the magnitude  of the exchange interactions~\cite{Katukuri_2014,Laurell:2020aa} read, $(\Delta_{\rm c}, A)\sim (0.35~\rm{eV},9~\rm{meV})$ for $\rm{Na}_2\rm{IrO}_3$ and $(\Delta_{\rm c}, A)\sim (1.1-1.9~\rm{eV}, 4~\rm{meV})$ for $\alpha$-$\rm{Ru}\rm{Cl}_3$.
As we will show below,  for model parameters corresponding to the zig-zag spin  ordering   observed in $\alpha$-$\rm{Ru}\rm{Cl}_3$   we can   reach temperature scales $2.6$ times lower than the exchange coupling, that is, 18K.  Hence  the overlap with temperature range where experimental results can be interpreted in terms of  Majorana fermions,  $ T \in [10,100 ]\rm{K} $,  is substantial \cite{Do:2017aa,Banerjee1055}. 
Henceforth, we  will consider  a $V=32$ lattice, which is beyond the accessible lattice size in exact diagonalization calculations (i.e., $V=24$ lattice)~\cite{Chaloupka2010,Chaloupka-2013, Katukuri_2014,Yamaji-1,Winter:2017aa,Winter2018,Laurell:2020aa}.
As for the Trotter discretization   we have used a range of  $\Delta\tau  \in [0.01, 0.1]$  depending upon the temperature.   For this range of  $\Delta\tau $ the systematic error is contained within our  error bars.  Values of $\beta U =10$  were found to be sufficient to guarantee projection to the odd  parity sector.

\begin{figure}
\centering
\centerline{\includegraphics[width=0.475\textwidth]{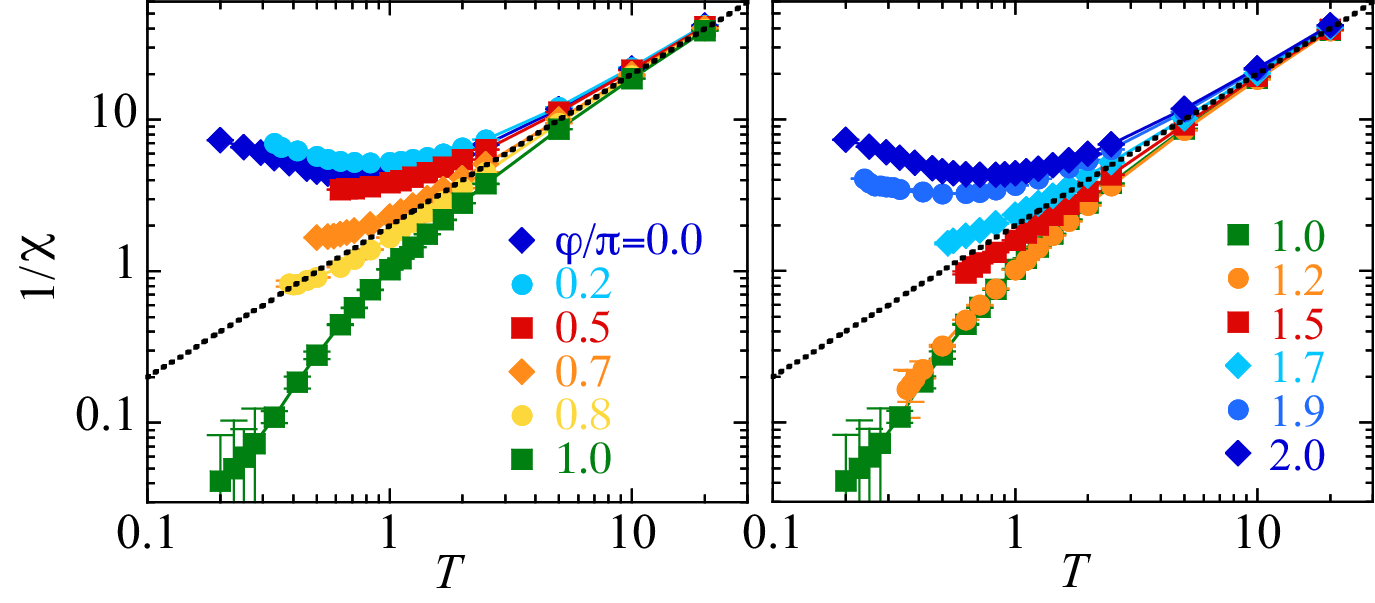}}
\caption{\label{fig:sus} 
$T$ dependence of inverse uniform spin susceptibilities $1/\chi$ at different values of $\varphi/\pi$.
Dashed line indicates the Curie's law considered here. 
}
\label{fig:unisus} 
\end{figure}

The ground-state phase diagram as a function of the angle $\varphi$ presented in Ref.~\cite{Chaloupka-2013} (see Fig.~\ref{fig:Fig1}(c)) reflects the competition between the isotropic Heisenberg exchange $J$ and the Kitaev-type bond-directional exchange $K$, and leads to antiferromagnetic (AFM), Kitaev spin liquid (KSL), zig-zag, ferromagnetic (FM), and stripy phases.
To study temperature effects as a function of $\varphi$, we measure the spin susceptibility,
\begin{eqnarray}
\label{eq:spinsus}
\chi_{\alpha}(\vec{q})=   \int_{0}^{\beta} \text{d} \tau 
\langle  \hat{\vec{O}}^{\alpha}_{\vec{q}}(\tau)  \hat{\vec{O}}^{\alpha}_{-\vec{q}}(0) \rangle, \nonumber\\ 
\end{eqnarray}
where $ \hat{\vec{O}}^{\alpha}_{\vec{q}}  = \frac{1}{\sqrt{V}} \sum_{\vec{r}} e^{i  \vec{q} \cdot \vec{r}} \left( \hat{S}_{\vec{r},A}^{\alpha}+\hat{S}_{\vec{r},B}^{\alpha}e^{i\vec{q}\vec{R}} \right)$.
Here $\ve{r}$ runs over the $A$ sublattice (or unit cell)  and $\vec{R}=2/3( \boldsymbol{a_2}-\boldsymbol{a_1}/2)$.

The uniform spin susceptibility  reads  $\chi=\frac{1}{3}\sum_{\alpha}\chi_{\alpha}(\vec{q=\Gamma})$ and
Fig.~\ref{fig:unisus} plots  $1/\chi$ for the various angles $\varphi$  down to the lowest accessible  temperature.  
In the  absence of  sign problem at $\varphi/\pi=0$ and $1$ we can access arbitrarily  low temperatures.
For all values of the angle $\varphi $, $\chi$ shows a Curie law at high temperatures.  The deviation from this law  marks an energy scale   that   allows for different interpretations.   One possibility is the onset of local  spin correlations.  In particular,  in the FM case, $\varphi/\pi = 1$, where $\vec{\Gamma}$  corresponds to the ordering wave vector   $\chi$  grows  and ultimately diverges at low temperatures. In contrast, in the AFM case, $\varphi/\pi = 0$, local antiferromagnetic correlations  lead to  a suppression of $\chi$  with respect the high temperature Curie law.  At low temperatures $\chi$ scales  to  a constant reflecting Goldstone modes.  In the absence of ordering, especially at angles  \textit{close}  to the Kitaev phases, 
the departure  from the Curie law  calls for different interpretations.   One possibility is that frustration effects lowers the temperature scale at which  local magnetic correlations develop. Other interpretations, put forward in Ref.~\cite{Do:2017aa},  argued in terms of  itinerant Majorana fermions akin to the Kitaev model~\cite{Kitaev06}.

 \begin{figure}
\centering
\centerline{\includegraphics[width=0.475\textwidth]{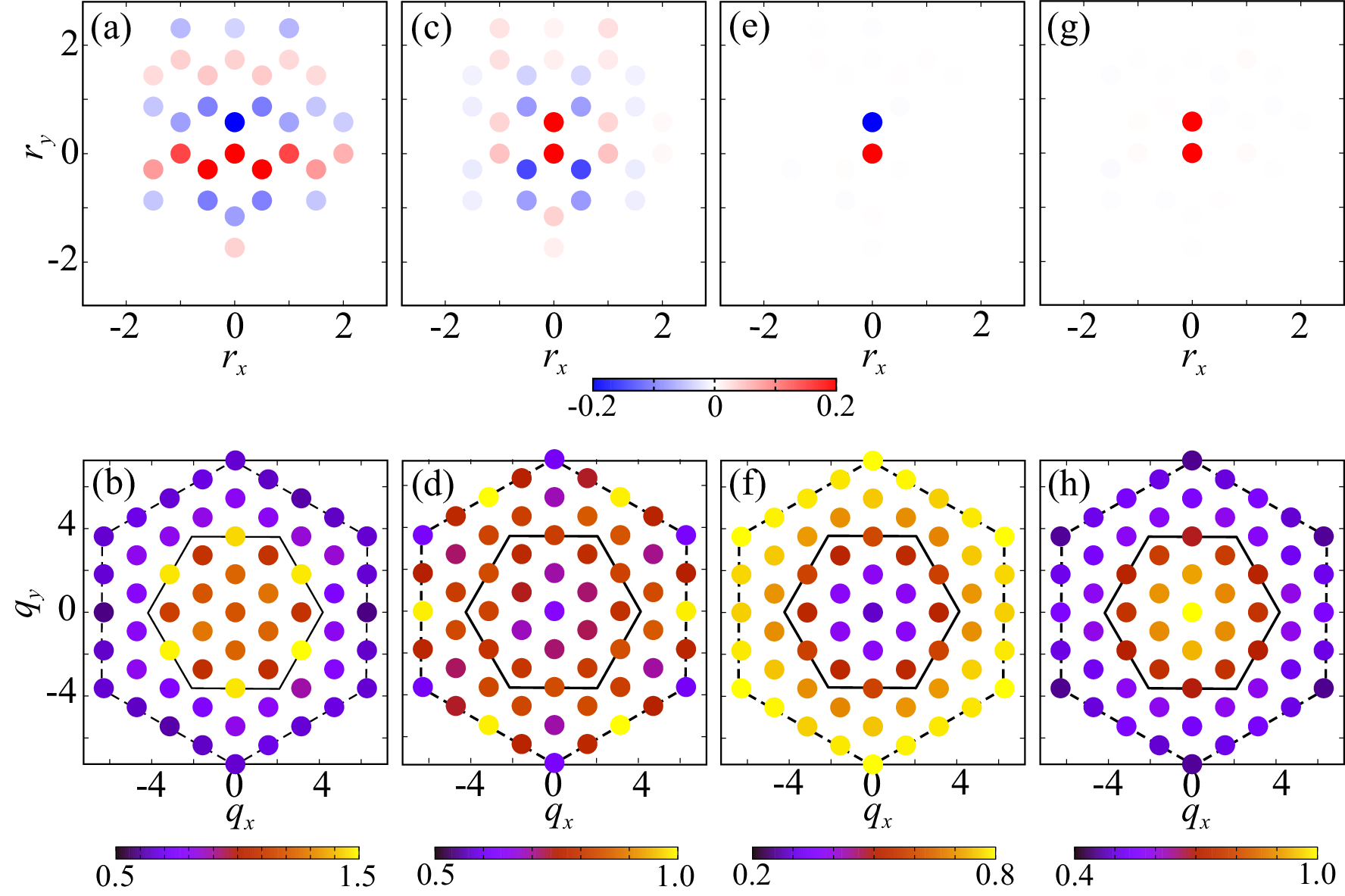}}
\caption{
Real-space spin-spin correlations $\langle  \hat{S}^{1}_{\vec{r}} \hat{S}^{1}_{\vec{0}} \rangle$ (top panel) and momentum  resolved spin susceptibility
$\chi(\vec{q})=\frac{1}{3}\sum_{\alpha}\chi_{\alpha}(\vec{q})$ (bottom panel)  in the  first (solid) and second (dashed line) Brillouin zones (see Fig.~\ref{fig:Fig1}(b)).
(a)-(b) $\varphi/\pi=0.8$ [$T=1/2.6$], (c)-(d) $\varphi/\pi=1.7$ [$T=1/1.9$], (e)-(f) $\varphi/\pi=0.5$ [$T=1/1.6$], and (g)-(h) $\varphi/\pi=1.5$ [$T=1/1.6$].
}
\label{fig:susq} 
\end{figure}

\begin{figure}
\centering
\centerline{\includegraphics[width=0.45\textwidth]{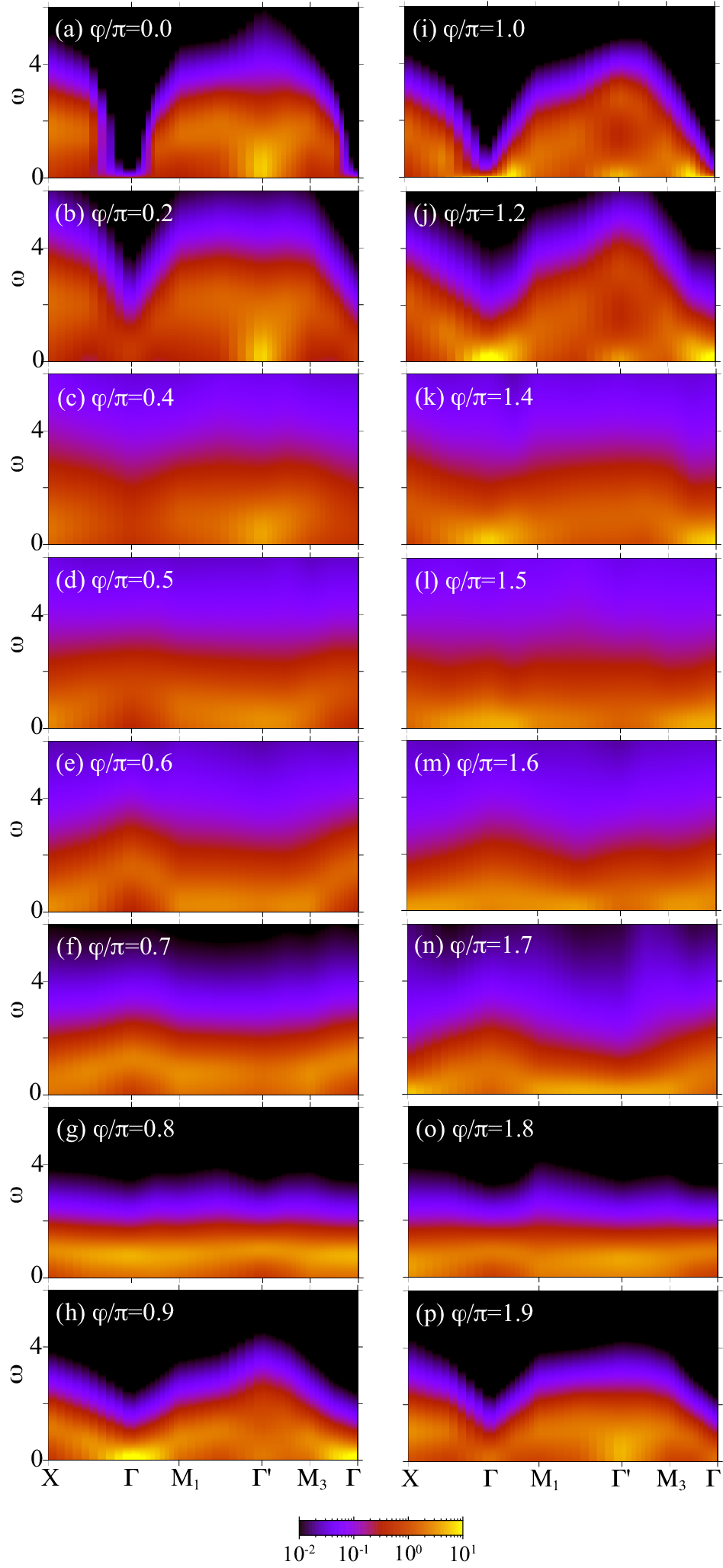}}
\caption{ Dynamical spin structure factor $C(\boldsymbol{q}, \omega)$ at different values of $\varphi/\pi$.
 Here, $T=1/1.6$.
 Results used here correspond to scans along the red 
 line of Fig.~\ref{fig:Fig1}(b).
}
\label{fig:Cqomega1}   
\end{figure}

We can confirm the above  by computing  real space spin-spin correlations in the zig-zag, stripy, and  Kitaev phases at temperatures    scales where  $\chi$ departs  from the Curie law. 
The zig-zag phase is characterized by  antiferromagnerically ordered,  ferromagnetic zig-zag rows of spins.   This ordering is apparent in  
$\langle  \hat{S}^{1}_{\vec{r}} \hat{S}^{1}_{\vec{0}} \rangle$ shown in Fig.~\ref{fig:susq}(a). 
The stripy phase  is characterized by antiferromagnerically ordered,  ferromagnetic  lines of spins.   This ordering is apparent in  Fig.~\ref{fig:susq}(c). 
On the other hand, in the antiferromagnetic (Fig.~\ref{fig:susq}(e))   and ferromagnetic (Fig.~\ref{fig:susq}(g))  Kitaev phases, real space spin correlations are limited to the nearest  neighbors. 
Fig.~\ref{fig:susq}   equally plots  the momentum  resolved spin susceptibility, $\chi(\vec{q})=\frac{1}{3}\sum_{\alpha}\chi_{\alpha}(\vec{q})$.     As apparent the zig-zag,  Fig.~\ref{fig:susq}(b), and stripy,  Fig.~\ref{fig:susq}(d), phases are characterized by   distinct 
precursors of  Bragg peaks.  On the other hand,  in the Kitaev limit only broad  features are apparent around the $\vec{\Gamma}~(\vec{\Gamma}'$) point for the FM (AFM) case.

\begin{figure}
\centering
\centerline{\includegraphics[width=0.45\textwidth]{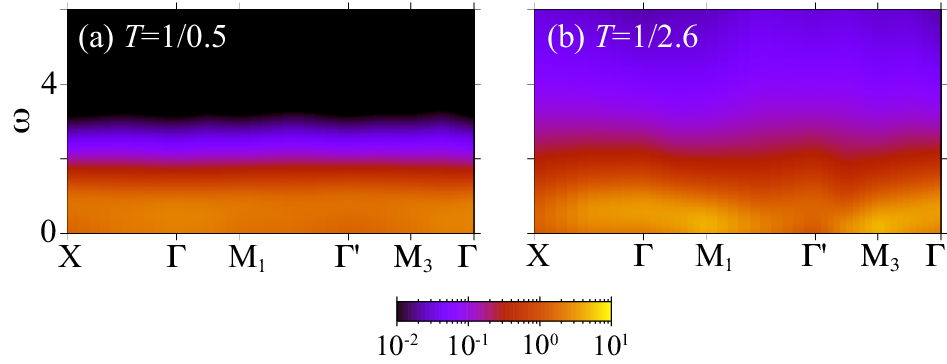}}
\caption{
Dynamical spin structure factor $C(\boldsymbol{q}, \omega)$ at $\varphi/\pi=0.8$ for higher [(a)$T=1/0.5$] and lower [(b)$T=1/2.6$] temperatures.
}
\label{fig:Cqomega2}   
\end{figure}

We now  turn our attention to the evolution of the dynamical spin structure factor as a function of  angle $\varphi$ and temperature. Such calculations are of experimental relevance  for the modeling of recent inelastic neutron scattering measurements~\cite{Choi2012, Banerjee1055,Do:2017aa}.
This quantity is defined as   $C(\boldsymbol{q}, \omega) =  \text{Im} \chi(\boldsymbol{q}, \omega) / \left( 1 - e^{-\beta \omega} \right) $ with 
\begin{eqnarray}
  \label{CSq}
\chi(\boldsymbol{q}, \omega)
&  &=\frac{i}{3} \sum_{\gamma}
 \int_0^{\infty} dt  \, 
 e^{i\omega t}
\left<   \left[ \hat{\boldsymbol{O}}^{\gamma}_{\boldsymbol{q}} , \hat{\boldsymbol{O}}^{\gamma}_{\boldsymbol{-q}}(-t) \right] \right>.
\end{eqnarray}
We compute this quantity  using the  stochastic  analytical continuation method \cite{Beach04a}. 
In the high temperature limit where we observe a Curie law  of the susceptibility  ($T > 10$),  we expect   $C(\boldsymbol{q}, \omega)$   to show no momentum dependence, and spectral  weight centered around $\omega \sim 0$.  Data at $T=10$ is shown in the Supplemental Material.    At $T = 1/1.6$  Fig.~\ref{fig:Cqomega1} shows  that the angle dependence of $C(\boldsymbol{q}, \omega)$  is pronounced and that the distinct  features of the ordered and disordered phases are apparent. 
For the KSL at $\varphi/\pi=0.5$, we see intensity located along the $\vec{\rm M_1}-\vec{\rm M_3}$ line as well as around the $\vec{\rm X}$ point.  In contrast, strong intensity around the $\vec{\Gamma}$ point is apparent in  the FM case ($\varphi/\pi=1.5$).
Similar behavior has been reported for the Kitaev model~\cite{Yoshioka2016} below a temperature scale related the coherence scale of the  Majorana fermions~\cite{Nasu2015}.   
In the Heisenberg limits, our data produce the well-known features  of  the  spin-wave dispersion relation: a quadratic dispersion around $\vec{\Gamma}$  the FM case ($\varphi/\pi=1$),  and a linear one around $\vec{\Gamma}'$ for the AFM  ($\varphi/\pi=0$).
Moving away form the AFM or FM phase towards the KSLs (Fig.~\ref{fig:Cqomega1}(a)-(d)  and  (i)-(l))  our data  shows the progressive vanishing of the spin wave  features.  
At $\varphi / \pi = 0.7$   we observe a buildup of low-lying spectral weight at the $\vec{\rm  M_1}$  and  $\vec{\rm  M_3}$ points as appropriate for  the zig-zag ordering. In contrast in the stripy phase at  $\varphi/\pi = 1.7$   we observe substantial low-lying weight at the $\vec{\rm  X}$  and $\vec{\Gamma}'$ points.   Note that this data is taken at higher temperatures than the susceptibility results of  Fig.~\ref{fig:susq} (c)  and (d) and that  $\varphi/\pi = 1.7$ is  close to the AFM phase. We understand the low-lying weight at  $\vec{\Gamma'} $  as a combined effect  of temperature and proximity to the AFM phase.  
   It  is  interesting to consider the temperature dependence of the zig-zag phase proximate to the KSL at $\varphi/\pi = 0.8$.   As a function of decreasing temperature,  Figs.~\ref{fig:Cqomega2}(a), \ref{fig:Cqomega1}(g) and \ref{fig:Cqomega2}(b)   we first observe a  buildup of weight around the $\vec{\Gamma}$  point followed by a softening at  $\vec{\rm  M_1}$ and $\vec{\rm  M_3}$.  The low temperature dynamical spin structure factor with high (low) energy weight at $\vec{\Gamma}$ ($\vec{\rm  M_1}$ and $\vec{\rm  M_3}$)  bears  similarities to inelastic neutron scattering experiments for $\alpha$-RuCl$_3$  reported in~\cite{Banerjee1055,Do:2017aa}.

{\it Summary and discussion.}---
We  have defined a formulation of the auxiliary field QMC algorithm for the generalized Kitaev model of Eq.~(\ref{Eq:KHM})   in which the imaginary part of the action is pinned by symmetry to  $0$ or $\pi$.  It turns out that this phase pinning strategy greatly improves the negative sign problem   and opens a window of temperatures  relevant to experiments where simulations can be carried out.  We  demonstrate this by carrying out extensive simulations of thermodynamical and dynamical properties of the Kitaev-Heisenberg model.
Aside from the magnetic susceptibility  and dynamical spin structure factor presented in this Letter, we can  compute the  specific heat, the  magnetotropic coefficient ~\cite{Modic20} as well as  heat transport~\cite{Kasahara2018}.
Furthermore our numerical method for the generalized Kitaev model of Eq.~(\ref{Eq:KHM}) can be applied to longer ranged interactions as well as off-diagonal $ \Gamma^{\alpha,\beta }$ interactions regarding specific materials such as $\rm{Na}_2\rm{IrO}_3$ and $\alpha$-$\rm{Ru}\rm{Cl}_3$.
Comparison of the aforementioned quantities with experimental data over a wide temperature range provides a useful tool to  determine model parameters.

\bigskip
\begin{acknowledgments}
We thank  K.  Modic and  R. Valenti  for motivating discussions.   
We also thank R. Thomale for bringing longer ranged interactions to our attention.
The authors gratefully acknowledge the Gauss Centre for Supercomputing e.V. (www.gauss-centre.eu) for funding this project by providing computing time on the GCS Supercomputer SUPERMUC-NG at the Leibniz Supercomputing Centre (www.lrz.de).
TS thanks funding from the Deutsche Forschungsgemeinschaft under the grant number SA 3986/1-1.
FFA   thanks financial support from the Deutsche Forschungsgemeinschaft,  Project C01 of the  SFB 1170,     as well as the W\"urzburg-Dresden Cluster of Excellence on Complexity and Topology in Quantum Matter ct.qmat (EXC 2147, project-id 390858490). 
\end{acknowledgments}

%\bibliography{./fassaad.bib,./toshihiro.bib,./refs.bib}

\begin{thebibliography}{41}%
\makeatletter
\providecommand \@ifxundefined [1]{%
 \@ifx{#1\undefined}
}%
\providecommand \@ifnum [1]{%
 \ifnum #1\expandafter \@firstoftwo
 \else \expandafter \@secondoftwo
 \fi
}%
\providecommand \@ifx [1]{%
 \ifx #1\expandafter \@firstoftwo
 \else \expandafter \@secondoftwo
 \fi
}%
\providecommand \natexlab [1]{#1}%
\providecommand \enquote  [1]{``#1''}%
\providecommand \bibnamefont  [1]{#1}%
\providecommand \bibfnamefont [1]{#1}%
\providecommand \citenamefont [1]{#1}%
\providecommand \href@noop [0]{\@secondoftwo}%
\providecommand \href [0]{\begingroup \@sanitize@url \@href}%
\providecommand \@href[1]{\@@startlink{#1}\@@href}%
\providecommand \@@href[1]{\endgroup#1\@@endlink}%
\providecommand \@sanitize@url [0]{\catcode `\\12\catcode `\$12\catcode
  `\&12\catcode `\#12\catcode `\^12\catcode `\_12\catcode `\%12\relax}%
\providecommand \@@startlink[1]{}%
\providecommand \@@endlink[0]{}%
\providecommand \url  [0]{\begingroup\@sanitize@url \@url }%
\providecommand \@url [1]{\endgroup\@href {#1}{\urlprefix }}%
\providecommand \urlprefix  [0]{URL }%
\providecommand \Eprint [0]{\href }%
\providecommand \doibase [0]{http://dx.doi.org/}%
\providecommand \selectlanguage [0]{\@gobble}%
\providecommand \bibinfo  [0]{\@secondoftwo}%
\providecommand \bibfield  [0]{\@secondoftwo}%
\providecommand \translation [1]{[#1]}%
\providecommand \BibitemOpen [0]{}%
\providecommand \bibitemStop [0]{}%
\providecommand \bibitemNoStop [0]{.\EOS\space}%
\providecommand \EOS [0]{\spacefactor3000\relax}%
\providecommand \BibitemShut  [1]{\csname bibitem#1\endcsname}%
\let\auto@bib@innerbib\@empty
%</preamble>
\bibitem [{\citenamefont {Witczak-Krempa}\ \emph {et~al.}(2014)\citenamefont
  {Witczak-Krempa}, \citenamefont {Chen}, \citenamefont {Kim},\ and\
  \citenamefont {Balents}}]{Kim_rev2013}%
  \BibitemOpen
  \bibfield  {author} {\bibinfo {author} {\bibfnamefont {W.}~\bibnamefont
  {Witczak-Krempa}}, \bibinfo {author} {\bibfnamefont {G.}~\bibnamefont
  {Chen}}, \bibinfo {author} {\bibfnamefont {Y.~B.}\ \bibnamefont {Kim}}, \
  and\ \bibinfo {author} {\bibfnamefont {L.}~\bibnamefont {Balents}},\ }\href
  {\doibase 10.1146/annurev-conmatphys-020911-125138} {\bibfield  {journal}
  {\bibinfo  {journal} {Annual Review of Condensed Matter Physics}\ }\textbf
  {\bibinfo {volume} {5}},\ \bibinfo {pages} {57} (\bibinfo {year}
  {2014})}\BibitemShut {NoStop}%
\bibitem [{\citenamefont {Jackeli}\ and\ \citenamefont
  {Khaliullin}(2009)}]{Jackeli-KSL-01}%
  \BibitemOpen
  \bibfield  {author} {\bibinfo {author} {\bibfnamefont {G.}~\bibnamefont
  {Jackeli}}\ and\ \bibinfo {author} {\bibfnamefont {G.}~\bibnamefont
  {Khaliullin}},\ }\href {\doibase 10.1103/PhysRevLett.102.017205} {\bibfield
  {journal} {\bibinfo  {journal} {Phys. Rev. Lett.}\ }\textbf {\bibinfo
  {volume} {102}},\ \bibinfo {pages} {017205} (\bibinfo {year}
  {2009})}\BibitemShut {NoStop}%
\bibitem [{\citenamefont {Chaloupka}\ \emph {et~al.}(2013)\citenamefont
  {Chaloupka}, \citenamefont {Jackeli},\ and\ \citenamefont
  {Khaliullin}}]{Chaloupka-2013}%
  \BibitemOpen
  \bibfield  {author} {\bibinfo {author} {\bibfnamefont {J.}~\bibnamefont
  {Chaloupka}}, \bibinfo {author} {\bibfnamefont {G.}~\bibnamefont {Jackeli}},
  \ and\ \bibinfo {author} {\bibfnamefont {G.}~\bibnamefont {Khaliullin}},\
  }\href {\doibase 10.1103/PhysRevLett.110.097204} {\bibfield  {journal}
  {\bibinfo  {journal} {Phys. Rev. Lett.}\ }\textbf {\bibinfo {volume} {110}},\
  \bibinfo {pages} {097204} (\bibinfo {year} {2013})}\BibitemShut {NoStop}%
\bibitem [{\citenamefont {Plumb}\ \emph {et~al.}(2014)\citenamefont {Plumb},
  \citenamefont {Clancy}, \citenamefont {Sandilands}, \citenamefont {Shankar},
  \citenamefont {Hu}, \citenamefont {Burch}, \citenamefont {Kee},\ and\
  \citenamefont {Kim}}]{Plumb2014}%
  \BibitemOpen
  \bibfield  {author} {\bibinfo {author} {\bibfnamefont {K.~W.}\ \bibnamefont
  {Plumb}}, \bibinfo {author} {\bibfnamefont {J.~P.}\ \bibnamefont {Clancy}},
  \bibinfo {author} {\bibfnamefont {L.~J.}\ \bibnamefont {Sandilands}},
  \bibinfo {author} {\bibfnamefont {V.~V.}\ \bibnamefont {Shankar}}, \bibinfo
  {author} {\bibfnamefont {Y.~F.}\ \bibnamefont {Hu}}, \bibinfo {author}
  {\bibfnamefont {K.~S.}\ \bibnamefont {Burch}}, \bibinfo {author}
  {\bibfnamefont {H.-Y.}\ \bibnamefont {Kee}}, \ and\ \bibinfo {author}
  {\bibfnamefont {Y.-J.}\ \bibnamefont {Kim}},\ }\href {\doibase
  10.1103/PhysRevB.90.041112} {\bibfield  {journal} {\bibinfo  {journal} {Phys.
  Rev. B}\ }\textbf {\bibinfo {volume} {90}},\ \bibinfo {pages} {041112(R)}
  (\bibinfo {year} {2014})}\BibitemShut {NoStop}%
\bibitem [{\citenamefont {Kitaev}(2006)}]{Kitaev06}%
  \BibitemOpen
  \bibfield  {author} {\bibinfo {author} {\bibfnamefont {A.}~\bibnamefont
  {Kitaev}},\ }\href {\doibase http://dx.doi.org/10.1016/j.aop.2005.10.005}
  {\bibfield  {journal} {\bibinfo  {journal} {Annals of Physics}\ }\textbf
  {\bibinfo {volume} {321}},\ \bibinfo {pages} {2 } (\bibinfo {year}
  {2006})}\BibitemShut {NoStop}%
\bibitem [{\citenamefont {Do}\ \emph {et~al.}(2017)\citenamefont {Do},
  \citenamefont {Park}, \citenamefont {Yoshitake}, \citenamefont {Nasu},
  \citenamefont {Motome}, \citenamefont {Kwon}, \citenamefont {Adroja},
  \citenamefont {Voneshen}, \citenamefont {Kim}, \citenamefont {Jang},
  \citenamefont {Park}, \citenamefont {Choi},\ and\ \citenamefont
  {Ji}}]{Do:2017aa}%
  \BibitemOpen
  \bibfield  {author} {\bibinfo {author} {\bibfnamefont {S.-H.}\ \bibnamefont
  {Do}}, \bibinfo {author} {\bibfnamefont {S.-Y.}\ \bibnamefont {Park}},
  \bibinfo {author} {\bibfnamefont {J.}~\bibnamefont {Yoshitake}}, \bibinfo
  {author} {\bibfnamefont {J.}~\bibnamefont {Nasu}}, \bibinfo {author}
  {\bibfnamefont {Y.}~\bibnamefont {Motome}}, \bibinfo {author} {\bibfnamefont
  {Y.~S.}\ \bibnamefont {Kwon}}, \bibinfo {author} {\bibfnamefont {D.~T.}\
  \bibnamefont {Adroja}}, \bibinfo {author} {\bibfnamefont {D.~J.}\
  \bibnamefont {Voneshen}}, \bibinfo {author} {\bibfnamefont {K.}~\bibnamefont
  {Kim}}, \bibinfo {author} {\bibfnamefont {T.~H.}\ \bibnamefont {Jang}},
  \bibinfo {author} {\bibfnamefont {J.~H.}\ \bibnamefont {Park}}, \bibinfo
  {author} {\bibfnamefont {K.-Y.}\ \bibnamefont {Choi}}, \ and\ \bibinfo
  {author} {\bibfnamefont {S.}~\bibnamefont {Ji}},\ }\href {\doibase
  10.1038/nphys4264} {\bibfield  {journal} {\bibinfo  {journal} {Nature
  Physics}\ }\textbf {\bibinfo {volume} {13}},\ \bibinfo {pages} {1079}
  (\bibinfo {year} {2017})}\BibitemShut {NoStop}%
\bibitem [{\citenamefont {Banerjee}\ \emph {et~al.}(2017)\citenamefont
  {Banerjee}, \citenamefont {Yan}, \citenamefont {Knolle}, \citenamefont
  {Bridges}, \citenamefont {Stone}, \citenamefont {Lumsden}, \citenamefont
  {Mandrus}, \citenamefont {Tennant}, \citenamefont {Moessner},\ and\
  \citenamefont {Nagler}}]{Banerjee1055}%
  \BibitemOpen
  \bibfield  {author} {\bibinfo {author} {\bibfnamefont {A.}~\bibnamefont
  {Banerjee}}, \bibinfo {author} {\bibfnamefont {J.}~\bibnamefont {Yan}},
  \bibinfo {author} {\bibfnamefont {J.}~\bibnamefont {Knolle}}, \bibinfo
  {author} {\bibfnamefont {C.~A.}\ \bibnamefont {Bridges}}, \bibinfo {author}
  {\bibfnamefont {M.~B.}\ \bibnamefont {Stone}}, \bibinfo {author}
  {\bibfnamefont {M.~D.}\ \bibnamefont {Lumsden}}, \bibinfo {author}
  {\bibfnamefont {D.~G.}\ \bibnamefont {Mandrus}}, \bibinfo {author}
  {\bibfnamefont {D.~A.}\ \bibnamefont {Tennant}}, \bibinfo {author}
  {\bibfnamefont {R.}~\bibnamefont {Moessner}}, \ and\ \bibinfo {author}
  {\bibfnamefont {S.~E.}\ \bibnamefont {Nagler}},\ }\href {\doibase
  10.1126/science.aah6015} {\bibfield  {journal} {\bibinfo  {journal}
  {Science}\ }\textbf {\bibinfo {volume} {356}},\ \bibinfo {pages} {1055}
  (\bibinfo {year} {2017})}\BibitemShut {NoStop}%
\bibitem [{\citenamefont {Chaloupka}\ \emph {et~al.}(2010)\citenamefont
  {Chaloupka}, \citenamefont {Jackeli},\ and\ \citenamefont
  {Khaliullin}}]{Chaloupka2010}%
  \BibitemOpen
  \bibfield  {author} {\bibinfo {author} {\bibfnamefont {J.}~\bibnamefont
  {Chaloupka}}, \bibinfo {author} {\bibfnamefont {G.}~\bibnamefont {Jackeli}},
  \ and\ \bibinfo {author} {\bibfnamefont {G.}~\bibnamefont {Khaliullin}},\
  }\href {\doibase 10.1103/PhysRevLett.105.027204} {\bibfield  {journal}
  {\bibinfo  {journal} {Phys. Rev. Lett.}\ }\textbf {\bibinfo {volume} {105}},\
  \bibinfo {pages} {027204} (\bibinfo {year} {2010})}\BibitemShut {NoStop}%
\bibitem [{\citenamefont {Katukuri}\ \emph {et~al.}(2014)\citenamefont
  {Katukuri}, \citenamefont {Nishimoto}, \citenamefont {Yushankhai},
  \citenamefont {Stoyanova}, \citenamefont {Kandpal}, \citenamefont {Choi},
  \citenamefont {Coldea}, \citenamefont {Rousochatzakis}, \citenamefont
  {Hozoi},\ and\ \citenamefont {van~den Brink}}]{Katukuri_2014}%
  \BibitemOpen
  \bibfield  {author} {\bibinfo {author} {\bibfnamefont {V.~M.}\ \bibnamefont
  {Katukuri}}, \bibinfo {author} {\bibfnamefont {S.}~\bibnamefont {Nishimoto}},
  \bibinfo {author} {\bibfnamefont {V.}~\bibnamefont {Yushankhai}}, \bibinfo
  {author} {\bibfnamefont {A.}~\bibnamefont {Stoyanova}}, \bibinfo {author}
  {\bibfnamefont {H.}~\bibnamefont {Kandpal}}, \bibinfo {author} {\bibfnamefont
  {S.}~\bibnamefont {Choi}}, \bibinfo {author} {\bibfnamefont {R.}~\bibnamefont
  {Coldea}}, \bibinfo {author} {\bibfnamefont {I.}~\bibnamefont
  {Rousochatzakis}}, \bibinfo {author} {\bibfnamefont {L.}~\bibnamefont
  {Hozoi}}, \ and\ \bibinfo {author} {\bibfnamefont {J.}~\bibnamefont {van~den
  Brink}},\ }\href {\doibase 10.1088/1367-2630/16/1/013056} {\bibfield
  {journal} {\bibinfo  {journal} {New Journal of Physics}\ }\textbf {\bibinfo
  {volume} {16}},\ \bibinfo {pages} {013056} (\bibinfo {year}
  {2014})}\BibitemShut {NoStop}%
\bibitem [{\citenamefont {Yamaji}\ \emph {et~al.}(2016)\citenamefont {Yamaji},
  \citenamefont {Suzuki}, \citenamefont {Yamada}, \citenamefont {Suga},
  \citenamefont {Kawashima},\ and\ \citenamefont {Imada}}]{Yamaji-1}%
  \BibitemOpen
  \bibfield  {author} {\bibinfo {author} {\bibfnamefont {Y.}~\bibnamefont
  {Yamaji}}, \bibinfo {author} {\bibfnamefont {T.}~\bibnamefont {Suzuki}},
  \bibinfo {author} {\bibfnamefont {T.}~\bibnamefont {Yamada}}, \bibinfo
  {author} {\bibfnamefont {S. I.}~\bibnamefont {Suga}}, \bibinfo {author}
  {\bibfnamefont {N.}~\bibnamefont {Kawashima}}, \ and\ \bibinfo {author}
  {\bibfnamefont {M.}~\bibnamefont {Imada}},\ }\href {\doibase
  10.1103/PhysRevB.93.174425} {\bibfield  {journal} {\bibinfo  {journal} {Phys.
  Rev. B}\ }\textbf {\bibinfo {volume} {93}},\ \bibinfo {pages} {174425}
  (\bibinfo {year} {2016})}\BibitemShut {NoStop}%
\bibitem [{\citenamefont {Winter}\ \emph
  {et~al.}(2017{\natexlab{a}})\citenamefont {Winter}, \citenamefont {Riedl},
  \citenamefont {Maksimov}, \citenamefont {Chernyshev}, \citenamefont
  {Honecker},\ and\ \citenamefont {Valent{\'\i}}}]{Winter:2017aa}%
  \BibitemOpen
  \bibfield  {author} {\bibinfo {author} {\bibfnamefont {S.~M.}\ \bibnamefont
  {Winter}}, \bibinfo {author} {\bibfnamefont {K.}~\bibnamefont {Riedl}},
  \bibinfo {author} {\bibfnamefont {P.~A.}\ \bibnamefont {Maksimov}}, \bibinfo
  {author} {\bibfnamefont {A.~L.}\ \bibnamefont {Chernyshev}}, \bibinfo
  {author} {\bibfnamefont {A.}~\bibnamefont {Honecker}}, \ and\ \bibinfo
  {author} {\bibfnamefont {R.}~\bibnamefont {Valent{\'\i}}},\ }\href {\doibase
  10.1038/s41467-017-01177-0} {\bibfield  {journal} {\bibinfo  {journal}
  {Nature Communications}\ }\textbf {\bibinfo {volume} {8}},\ \bibinfo {pages}
  {1152} (\bibinfo {year} {2017}{\natexlab{a}})}\BibitemShut {NoStop}%
\bibitem [{\citenamefont {Winter}\ \emph {et~al.}(2018)\citenamefont {Winter},
  \citenamefont {Riedl}, \citenamefont {Kaib}, \citenamefont {Coldea},\ and\
  \citenamefont {Valent\'{\i}}}]{Winter2018}%
  \BibitemOpen
  \bibfield  {author} {\bibinfo {author} {\bibfnamefont {S.~M.}\ \bibnamefont
  {Winter}}, \bibinfo {author} {\bibfnamefont {K.}~\bibnamefont {Riedl}},
  \bibinfo {author} {\bibfnamefont {D.}~\bibnamefont {Kaib}}, \bibinfo {author}
  {\bibfnamefont {R.}~\bibnamefont {Coldea}}, \ and\ \bibinfo {author}
  {\bibfnamefont {R.}~\bibnamefont {Valent\'{\i}}},\ }\href {\doibase
  10.1103/PhysRevLett.120.077203} {\bibfield  {journal} {\bibinfo  {journal}
  {Phys. Rev. Lett.}\ }\textbf {\bibinfo {volume} {120}},\ \bibinfo {pages}
  {077203} (\bibinfo {year} {2018})}\BibitemShut {NoStop}%
\bibitem [{\citenamefont {Laurell}\ and\ \citenamefont
  {Okamoto}(2020)}]{Laurell:2020aa}%
  \BibitemOpen
  \bibfield  {author} {\bibinfo {author} {\bibfnamefont {P.}~\bibnamefont
  {Laurell}}\ and\ \bibinfo {author} {\bibfnamefont {S.}~\bibnamefont
  {Okamoto}},\ }\href {\doibase 10.1038/s41535-019-0203-y} {\bibfield
  {journal} {\bibinfo  {journal} {npj Quantum Materials}\ }\textbf {\bibinfo
  {volume} {5}},\ \bibinfo {pages} {2} (\bibinfo {year} {2020})}\BibitemShut
  {NoStop}%
\bibitem [{\citenamefont {Reuther}\ \emph {et~al.}(2011)\citenamefont
  {Reuther}, \citenamefont {Thomale},\ and\ \citenamefont
  {Trebst}}]{Reuther2011}%
  \BibitemOpen
  \bibfield  {author} {\bibinfo {author} {\bibfnamefont {J.}~\bibnamefont
  {Reuther}}, \bibinfo {author} {\bibfnamefont {R.}~\bibnamefont {Thomale}}, \
  and\ \bibinfo {author} {\bibfnamefont {S.}~\bibnamefont {Trebst}},\ }\href
  {\doibase 10.1103/PhysRevB.84.100406} {\bibfield  {journal} {\bibinfo
  {journal} {Phys. Rev. B}\ }\textbf {\bibinfo {volume} {84}},\ \bibinfo
  {pages} {100406(R)} (\bibinfo {year} {2011})}\BibitemShut {NoStop}%
\bibitem [{\citenamefont {Singh}\ \emph {et~al.}(2012)\citenamefont {Singh},
  \citenamefont {Manni}, \citenamefont {Reuther}, \citenamefont {Berlijn},
  \citenamefont {Thomale}, \citenamefont {Ku}, \citenamefont {Trebst},\ and\
  \citenamefont {Gegenwart}}]{Singh2012}%
  \BibitemOpen
  \bibfield  {author} {\bibinfo {author} {\bibfnamefont {Y.}~\bibnamefont
  {Singh}}, \bibinfo {author} {\bibfnamefont {S.}~\bibnamefont {Manni}},
  \bibinfo {author} {\bibfnamefont {J.}~\bibnamefont {Reuther}}, \bibinfo
  {author} {\bibfnamefont {T.}~\bibnamefont {Berlijn}}, \bibinfo {author}
  {\bibfnamefont {R.}~\bibnamefont {Thomale}}, \bibinfo {author} {\bibfnamefont
  {W.}~\bibnamefont {Ku}}, \bibinfo {author} {\bibfnamefont {S.}~\bibnamefont
  {Trebst}}, \ and\ \bibinfo {author} {\bibfnamefont {P.}~\bibnamefont
  {Gegenwart}},\ }\href {\doibase 10.1103/PhysRevLett.108.127203} {\bibfield
  {journal} {\bibinfo  {journal} {Phys. Rev. Lett.}\ }\textbf {\bibinfo
  {volume} {108}},\ \bibinfo {pages} {127203} (\bibinfo {year}
  {2012})}\BibitemShut {NoStop}%
\bibitem [{\citenamefont {Gohlke}\ \emph {et~al.}(2017)\citenamefont {Gohlke},
  \citenamefont {Verresen}, \citenamefont {Moessner},\ and\ \citenamefont
  {Pollmann}}]{Gohlke-2017}%
  \BibitemOpen
  \bibfield  {author} {\bibinfo {author} {\bibfnamefont {M.}~\bibnamefont
  {Gohlke}}, \bibinfo {author} {\bibfnamefont {R.}~\bibnamefont {Verresen}},
  \bibinfo {author} {\bibfnamefont {R.}~\bibnamefont {Moessner}}, \ and\
  \bibinfo {author} {\bibfnamefont {F.}~\bibnamefont {Pollmann}},\ }\href
  {\doibase 10.1103/PhysRevLett.119.157203} {\bibfield  {journal} {\bibinfo
  {journal} {Phys. Rev. Lett.}\ }\textbf {\bibinfo {volume} {119}},\ \bibinfo
  {pages} {157203} (\bibinfo {year} {2017})}\BibitemShut {NoStop}%
\bibitem [{\citenamefont {Wan}\ \emph {et~al.}(2020)\citenamefont {Wan},
  \citenamefont {Zhang},\ and\ \citenamefont {Yao}}]{Wan20}%
  \BibitemOpen
  \bibfield  {author} {\bibinfo {author} {\bibfnamefont {Z.-Q.}\ \bibnamefont
  {Wan}}, \bibinfo {author} {\bibfnamefont {S.-X.}\ \bibnamefont {Zhang}}, \
  and\ \bibinfo {author} {\bibfnamefont {H.}~\bibnamefont {Yao}},\ }\href@noop
  {} {\bibfield  {journal} {\bibinfo  {journal} {arXiv:2010.01141}\ } (\bibinfo
  {year} {2020})},\ \Eprint {http://arxiv.org/abs/2010.01141} {arXiv:2010.01141
  [cond-mat.str-el]} \BibitemShut {NoStop}%
\bibitem [{\citenamefont {Hangleiter}\ \emph {et~al.}(2020)\citenamefont
  {Hangleiter}, \citenamefont {Roth}, \citenamefont {Nagaj},\ and\
  \citenamefont {Eisert}}]{Hangleiter20}%
  \BibitemOpen
  \bibfield  {author} {\bibinfo {author} {\bibfnamefont {D.}~\bibnamefont
  {Hangleiter}}, \bibinfo {author} {\bibfnamefont {I.}~\bibnamefont {Roth}},
  \bibinfo {author} {\bibfnamefont {D.}~\bibnamefont {Nagaj}}, \ and\ \bibinfo
  {author} {\bibfnamefont {J.}~\bibnamefont {Eisert}},\ }\href
  {https://advances.sciencemag.org/content/6/33/eabb8341} {\bibfield  {journal}
  {\bibinfo  {journal} {Science Advances}\ }\textbf {\bibinfo {volume} {6}}
  (\bibinfo {year} {2020})}\BibitemShut {NoStop}%
\bibitem [{\citenamefont {Blankenbecler}\ \emph {et~al.}(1981)\citenamefont
  {Blankenbecler}, \citenamefont {Scalapino},\ and\ \citenamefont
  {Sugar}}]{Blankenbecler81}%
  \BibitemOpen
  \bibfield  {author} {\bibinfo {author} {\bibfnamefont {R.}~\bibnamefont
  {Blankenbecler}}, \bibinfo {author} {\bibfnamefont {D.~J.}\ \bibnamefont
  {Scalapino}}, \ and\ \bibinfo {author} {\bibfnamefont {R.~L.}\ \bibnamefont
  {Sugar}},\ }\href {\doibase 10.1103/PhysRevD.24.2278} {\bibfield  {journal}
  {\bibinfo  {journal} {Phys. Rev. D}\ }\textbf {\bibinfo {volume} {24}},\
  \bibinfo {pages} {2278} (\bibinfo {year} {1981})}\BibitemShut {NoStop}%
\bibitem [{\citenamefont {White}\ \emph {et~al.}(1989)\citenamefont {White},
  \citenamefont {Scalapino}, \citenamefont {Sugar}, \citenamefont {Loh},
  \citenamefont {Gubernatis},\ and\ \citenamefont {Scalettar}}]{White89}%
  \BibitemOpen
  \bibfield  {author} {\bibinfo {author} {\bibfnamefont {S. R.}~\bibnamefont
  {White}}, \bibinfo {author} {\bibfnamefont {D. J.}~\bibnamefont {Scalapino}},
  \bibinfo {author} {\bibfnamefont {R. L.}~\bibnamefont {Sugar}}, \bibinfo
  {author} {\bibfnamefont {E. Y.}~\bibnamefont {Loh}}, \bibinfo {author}
  {\bibfnamefont {J. E.}~\bibnamefont {Gubernatis}}, \ and\ \bibinfo {author}
  {\bibfnamefont {R. T.}~\bibnamefont {Scalettar}},\ }\href {\doibase
  10.1103/PhysRevB.40.506} {\bibfield  {journal} {\bibinfo  {journal} {Phys.
  Rev. B}\ }\textbf {\bibinfo {volume} {40}},\ \bibinfo {pages} {506} (\bibinfo
  {year} {1989})}\BibitemShut {NoStop}%
\bibitem [{\citenamefont {Bercx}\ \emph {et~al.}(2017)\citenamefont {Bercx},
  \citenamefont {Goth}, \citenamefont {Hofmann},\ and\ \citenamefont
  {Assaad}}]{ALF_v1}%
  \BibitemOpen
  \bibfield  {author} {\bibinfo {author} {\bibfnamefont {M.}~\bibnamefont
  {Bercx}}, \bibinfo {author} {\bibfnamefont {F.}~\bibnamefont {Goth}},
  \bibinfo {author} {\bibfnamefont {J.~S.}\ \bibnamefont {Hofmann}}, \ and\
  \bibinfo {author} {\bibfnamefont {F.~F.}\ \bibnamefont {Assaad}},\ }\href
  {\doibase 10.21468/SciPostPhys.3.2.013} {\bibfield  {journal} {\bibinfo
  {journal} {SciPost Phys.}\ }\textbf {\bibinfo {volume} {3}},\ \bibinfo
  {pages} {013} (\bibinfo {year} {2017})}\BibitemShut {NoStop}%
\bibitem [{\citenamefont {Wu}\ and\ \citenamefont {Zhang}(2005)}]{Wu04}%
  \BibitemOpen
  \bibfield  {author} {\bibinfo {author} {\bibfnamefont {C.}~\bibnamefont
  {Wu}}\ and\ \bibinfo {author} {\bibfnamefont {S.-C.}\ \bibnamefont {Zhang}},\
  }\href {\doibase 10.1103/PhysRevB.71.155115} {\bibfield  {journal} {\bibinfo
  {journal} {Phys. Rev. B}\ }\textbf {\bibinfo {volume} {71}},\ \bibinfo
  {pages} {155115} (\bibinfo {year} {2005})}\BibitemShut {NoStop}%
\bibitem [{\citenamefont {Wei}\ \emph {et~al.}(2016)\citenamefont {Wei},
  \citenamefont {Wu}, \citenamefont {Li}, \citenamefont {Zhang},\ and\
  \citenamefont {Xiang}}]{Wei16}%
  \BibitemOpen
  \bibfield  {author} {\bibinfo {author} {\bibfnamefont {Z.~C.}\ \bibnamefont
  {Wei}}, \bibinfo {author} {\bibfnamefont {C.}~\bibnamefont {Wu}}, \bibinfo
  {author} {\bibfnamefont {Y.}~\bibnamefont {Li}}, \bibinfo {author}
  {\bibfnamefont {S.}~\bibnamefont {Zhang}}, \ and\ \bibinfo {author}
  {\bibfnamefont {T.}~\bibnamefont {Xiang}},\ }\href {\doibase
  10.1103/PhysRevLett.116.250601} {\bibfield  {journal} {\bibinfo  {journal}
  {Phys. Rev. Lett.}\ }\textbf {\bibinfo {volume} {116}},\ \bibinfo {pages}
  {250601} (\bibinfo {year} {2016})}\BibitemShut {NoStop}%
\bibitem [{\citenamefont {Li}\ \emph {et~al.}(2016)\citenamefont {Li},
  \citenamefont {Jiang},\ and\ \citenamefont {Yao}}]{Li16}%
  \BibitemOpen
  \bibfield  {author} {\bibinfo {author} {\bibfnamefont {Z.-X.}\ \bibnamefont
  {Li}}, \bibinfo {author} {\bibfnamefont {Y.-F.}\ \bibnamefont {Jiang}}, \
  and\ \bibinfo {author} {\bibfnamefont {H.}~\bibnamefont {Yao}},\ }\href
  {\doibase 10.1103/PhysRevLett.117.267002} {\bibfield  {journal} {\bibinfo
  {journal} {Phys. Rev. Lett.}\ }\textbf {\bibinfo {volume} {117}},\ \bibinfo
  {pages} {267002} (\bibinfo {year} {2016})}\BibitemShut {NoStop}%
\bibitem [{\citenamefont {Sato}\ \emph {et~al.}(2018)\citenamefont {Sato},
  \citenamefont {Assaad},\ and\ \citenamefont {Grover}}]{SatoT17_1}%
  \BibitemOpen
  \bibfield  {author} {\bibinfo {author} {\bibfnamefont {T.}~\bibnamefont
  {Sato}}, \bibinfo {author} {\bibfnamefont {F.~F.}\ \bibnamefont {Assaad}}, \
  and\ \bibinfo {author} {\bibfnamefont {T.}~\bibnamefont {Grover}},\ }\href
  {\doibase 10.1103/PhysRevLett.120.107201} {\bibfield  {journal} {\bibinfo
  {journal} {Phys. Rev. Lett.}\ }\textbf {\bibinfo {volume} {120}},\ \bibinfo
  {pages} {107201} (\bibinfo {year} {2018})}\BibitemShut {NoStop}%
\bibitem [{\citenamefont {Liu}\ \emph {et~al.}(2019)\citenamefont {Liu},
  \citenamefont {Wang}, \citenamefont {Sato}, \citenamefont {Hohenadler},
  \citenamefont {Wang}, \citenamefont {Guo},\ and\ \citenamefont
  {Assaad}}]{Liu18}%
  \BibitemOpen
  \bibfield  {author} {\bibinfo {author} {\bibfnamefont {Y.}~\bibnamefont
  {Liu}}, \bibinfo {author} {\bibfnamefont {Z.}~\bibnamefont {Wang}}, \bibinfo
  {author} {\bibfnamefont {T.}~\bibnamefont {Sato}}, \bibinfo {author}
  {\bibfnamefont {M.}~\bibnamefont {Hohenadler}}, \bibinfo {author}
  {\bibfnamefont {C.}~\bibnamefont {Wang}}, \bibinfo {author} {\bibfnamefont
  {W.}~\bibnamefont {Guo}}, \ and\ \bibinfo {author} {\bibfnamefont {F.~F.}\
  \bibnamefont {Assaad}},\ }\href {\doibase 10.1038/s41467-019-10372-0}
  {\bibfield  {journal} {\bibinfo  {journal} {Nature Communications}\ }\textbf
  {\bibinfo {volume} {10}},\ \bibinfo {pages} {2658} (\bibinfo {year}
  {2019})}\BibitemShut {NoStop}%
\bibitem [{\citenamefont {Ippoliti}\ \emph {et~al.}(2018)\citenamefont
  {Ippoliti}, \citenamefont {Mong}, \citenamefont {Assaad},\ and\ \citenamefont
  {Zaletel}}]{Ippoliti18}%
  \BibitemOpen
  \bibfield  {author} {\bibinfo {author} {\bibfnamefont {M.}~\bibnamefont
  {Ippoliti}}, \bibinfo {author} {\bibfnamefont {R.~S.~K.}\ \bibnamefont
  {Mong}}, \bibinfo {author} {\bibfnamefont {F.~F.}\ \bibnamefont {Assaad}}, \
  and\ \bibinfo {author} {\bibfnamefont {M.~P.}\ \bibnamefont {Zaletel}},\
  }\href {\doibase 10.1103/PhysRevB.98.235108} {\bibfield  {journal} {\bibinfo
  {journal} {Phys. Rev. B}\ }\textbf {\bibinfo {volume} {98}},\ \bibinfo
  {pages} {235108} (\bibinfo {year} {2018})}\BibitemShut {NoStop}%
\bibitem [{\citenamefont {Wang}\ \emph {et~al.}(2020)\citenamefont {Wang},
  \citenamefont {Zaletel}, \citenamefont {Mong},\ and\ \citenamefont
  {Assaad}}]{WangZ20}%
  \BibitemOpen
  \bibfield  {author} {\bibinfo {author} {\bibfnamefont {Z.}~\bibnamefont
  {Wang}}, \bibinfo {author} {\bibfnamefont {M.~P.}\ \bibnamefont {Zaletel}},
  \bibinfo {author} {\bibfnamefont {R.~S.~K.}\ \bibnamefont {Mong}}, \ and\
  \bibinfo {author} {\bibfnamefont {F.~F.}\ \bibnamefont {Assaad}},\
  }\href@noop {} {\bibfield  {journal} {\bibinfo  {journal} {arXiv:2003.08368}\
  } (\bibinfo {year} {2020})},\ \Eprint {http://arxiv.org/abs/2003.08368}
  {arXiv:2003.08368 [cond-mat.str-el]} \BibitemShut {NoStop}%
\bibitem [{\citenamefont {Schattner}\ \emph {et~al.}(2016)\citenamefont
  {Schattner}, \citenamefont {Lederer}, \citenamefont {Kivelson},\ and\
  \citenamefont {Berg}}]{Schattner15}%
  \BibitemOpen
  \bibfield  {author} {\bibinfo {author} {\bibfnamefont {Y.}~\bibnamefont
  {Schattner}}, \bibinfo {author} {\bibfnamefont {S.}~\bibnamefont {Lederer}},
  \bibinfo {author} {\bibfnamefont {S.~A.}\ \bibnamefont {Kivelson}}, \ and\
  \bibinfo {author} {\bibfnamefont {E.}~\bibnamefont {Berg}},\ }\href {\doibase
  10.1103/PhysRevX.6.031028} {\bibfield  {journal} {\bibinfo  {journal} {Phys.
  Rev. X}\ }\textbf {\bibinfo {volume} {6}},\ \bibinfo {pages} {031028}
  (\bibinfo {year} {2016})}\BibitemShut {NoStop}%
\bibitem [{\citenamefont {Pan}\ \emph {et~al.}(2020)\citenamefont {Pan},
  \citenamefont {Wang}, \citenamefont {Davis}, \citenamefont {Wang},\ and\
  \citenamefont {Meng}}]{Pan20}%
  \BibitemOpen
  \bibfield  {author} {\bibinfo {author} {\bibfnamefont {G.}~\bibnamefont
  {Pan}}, \bibinfo {author} {\bibfnamefont {W.}~\bibnamefont {Wang}}, \bibinfo
  {author} {\bibfnamefont {A.}~\bibnamefont {Davis}}, \bibinfo {author}
  {\bibfnamefont {Y.}~\bibnamefont {Wang}}, \ and\ \bibinfo {author}
  {\bibfnamefont {Z.~Y.}\ \bibnamefont {Meng}},\ }\href@noop {} {\bibfield
  {journal} {\bibinfo  {journal} {arXiv:2001.06586}\ } (\bibinfo {year}
  {2020})},\ \Eprint {http://arxiv.org/abs/2001.06586} {arXiv:2001.06586
  [cond-mat.str-el]} \BibitemShut {NoStop}%
\bibitem [{\citenamefont {Huffman}\ and\ \citenamefont
  {Chandrasekharan}(2014)}]{Huffman14}%
  \BibitemOpen
  \bibfield  {author} {\bibinfo {author} {\bibfnamefont {E.~F.}\ \bibnamefont
  {Huffman}}\ and\ \bibinfo {author} {\bibfnamefont {S.}~\bibnamefont
  {Chandrasekharan}},\ }\href {\doibase 10.1103/PhysRevB.89.111101} {\bibfield
  {journal} {\bibinfo  {journal} {Phys. Rev. B}\ }\textbf {\bibinfo {volume}
  {89}},\ \bibinfo {pages} {111101(R)} (\bibinfo {year} {2014})}\BibitemShut
  {NoStop}%
\bibitem [{\citenamefont {Huang}\ \emph {et~al.}(2018)\citenamefont {Huang},
  \citenamefont {Mendl}, \citenamefont {Jiang}, \citenamefont {Moritz},\ and\
  \citenamefont {Devereaux}}]{Huang18}%
  \BibitemOpen
  \bibfield  {author} {\bibinfo {author} {\bibfnamefont {E.~W.}\ \bibnamefont
  {Huang}}, \bibinfo {author} {\bibfnamefont {C.~B.}\ \bibnamefont {Mendl}},
  \bibinfo {author} {\bibfnamefont {H.-C.}\ \bibnamefont {Jiang}}, \bibinfo
  {author} {\bibfnamefont {B.}~\bibnamefont {Moritz}}, \ and\ \bibinfo {author}
  {\bibfnamefont {T.~P.}\ \bibnamefont {Devereaux}},\ }\href {\doibase
  10.1038/s41535-018-0097-0} {\bibfield  {journal} {\bibinfo  {journal} {npj
  Quantum Materials}\ }\textbf {\bibinfo {volume} {3}},\ \bibinfo {pages} {22}
  (\bibinfo {year} {2018})}\BibitemShut {NoStop}%
\bibitem [{\citenamefont {{Huang}}\ \emph {et~al.}(2018)\citenamefont
  {{Huang}}, \citenamefont {{Sheppard}}, \citenamefont {{Moritz}},\ and\
  \citenamefont {{Devereaux}}}]{Huang19}%
  \BibitemOpen
  \bibfield  {author} {\bibinfo {author} {\bibfnamefont {E.~W.}\ \bibnamefont
  {{Huang}}}, \bibinfo {author} {\bibfnamefont {R.}~\bibnamefont {{Sheppard}}},
  \bibinfo {author} {\bibfnamefont {B.}~\bibnamefont {{Moritz}}}, \ and\
  \bibinfo {author} {\bibfnamefont {T.~P.}\ \bibnamefont {{Devereaux}}},\
  }\href@noop {} {\bibfield  {journal} {\bibinfo  {journal} {arXiv e-prints}\
  ,\ \bibinfo {eid} {arXiv:1806.08346}} (\bibinfo {year} {2018})},\ \Eprint
  {http://arxiv.org/abs/1806.08346} {arXiv:1806.08346 [cond-mat.str-el]}
  \BibitemShut {NoStop}%
\bibitem [{\citenamefont {Assaad}\ and\ \citenamefont
  {Evertz}(2008)}]{Assaad08_rev}%
  \BibitemOpen
  \bibfield  {author} {\bibinfo {author} {\bibfnamefont {F.}~\bibnamefont
  {Assaad}}\ and\ \bibinfo {author} {\bibfnamefont {H.}~\bibnamefont
  {Evertz}},\ }in\ \href {\doibase 10.1007/978-3-540-74686-7_10} {\emph
  {\bibinfo {booktitle} {Computational Many-Particle Physics}}},\ \bibinfo
  {series} {Lecture Notes in Physics}, Vol.\ \bibinfo {volume} {739},\ \bibinfo
  {editor} {edited by\ \bibinfo {editor} {\bibfnamefont {H.}~\bibnamefont
  {Fehske}}, \bibinfo {editor} {\bibfnamefont {R.}~\bibnamefont {Schneider}}, \
  and\ \bibinfo {editor} {\bibfnamefont {A.}~\bibnamefont {Wei{\ss}e}}}\
  (\bibinfo  {publisher} {Springer},\ \bibinfo {address} {Berlin Heidelberg},\
  \bibinfo {year} {2008})\ pp.\ \bibinfo {pages} {277--356}\BibitemShut
  {NoStop}%
\bibitem [{\citenamefont {Winter}\ \emph
  {et~al.}(2017{\natexlab{b}})\citenamefont {Winter}, \citenamefont {Tsirlin},
  \citenamefont {Daghofer}, \citenamefont {van~den Brink}, \citenamefont
  {Singh}, \citenamefont {Gegenwart},\ and\ \citenamefont
  {Valent{\'{\i}}}}]{Winter_2017}%
  \BibitemOpen
  \bibfield  {author} {\bibinfo {author} {\bibfnamefont {S.~M.}\ \bibnamefont
  {Winter}}, \bibinfo {author} {\bibfnamefont {A.~A.}\ \bibnamefont {Tsirlin}},
  \bibinfo {author} {\bibfnamefont {M.}~\bibnamefont {Daghofer}}, \bibinfo
  {author} {\bibfnamefont {J.}~\bibnamefont {van~den Brink}}, \bibinfo {author}
  {\bibfnamefont {Y.}~\bibnamefont {Singh}}, \bibinfo {author} {\bibfnamefont
  {P.}~\bibnamefont {Gegenwart}}, \ and\ \bibinfo {author} {\bibfnamefont
  {R.}~\bibnamefont {Valent{\'{\i}}}},\ }\href {\doibase
  10.1088/1361-648x/aa8cf5} {\bibfield  {journal} {\bibinfo  {journal} {Journal
  of Physics: Condensed Matter}\ }\textbf {\bibinfo {volume} {29}},\ \bibinfo
  {pages} {493002} (\bibinfo {year} {2017}{\natexlab{b}})}\BibitemShut
  {NoStop}%
\bibitem [{\citenamefont {Choi}\ \emph {et~al.}(2012)\citenamefont {Choi},
  \citenamefont {Coldea}, \citenamefont {Kolmogorov}, \citenamefont
  {Lancaster}, \citenamefont {Mazin}, \citenamefont {Blundell}, \citenamefont
  {Radaelli}, \citenamefont {Singh}, \citenamefont {Gegenwart}, \citenamefont
  {Choi}, \citenamefont {Cheong}, \citenamefont {Baker}, \citenamefont
  {Stock},\ and\ \citenamefont {Taylor}}]{Choi2012}%
  \BibitemOpen
  \bibfield  {author} {\bibinfo {author} {\bibfnamefont {S.~K.}\ \bibnamefont
  {Choi}}, \bibinfo {author} {\bibfnamefont {R.}~\bibnamefont {Coldea}},
  \bibinfo {author} {\bibfnamefont {A.~N.}\ \bibnamefont {Kolmogorov}},
  \bibinfo {author} {\bibfnamefont {T.}~\bibnamefont {Lancaster}}, \bibinfo
  {author} {\bibfnamefont {I.~I.}\ \bibnamefont {Mazin}}, \bibinfo {author}
  {\bibfnamefont {S.~J.}\ \bibnamefont {Blundell}}, \bibinfo {author}
  {\bibfnamefont {P.~G.}\ \bibnamefont {Radaelli}}, \bibinfo {author}
  {\bibfnamefont {Y.}~\bibnamefont {Singh}}, \bibinfo {author} {\bibfnamefont
  {P.}~\bibnamefont {Gegenwart}}, \bibinfo {author} {\bibfnamefont {K.~R.}\
  \bibnamefont {Choi}}, \bibinfo {author} {\bibfnamefont {S.-W.}\ \bibnamefont
  {Cheong}}, \bibinfo {author} {\bibfnamefont {P.~J.}\ \bibnamefont {Baker}},
  \bibinfo {author} {\bibfnamefont {C.}~\bibnamefont {Stock}}, \ and\ \bibinfo
  {author} {\bibfnamefont {J.}~\bibnamefont {Taylor}},\ }\href {\doibase
  10.1103/PhysRevLett.108.127204} {\bibfield  {journal} {\bibinfo  {journal}
  {Phys. Rev. Lett.}\ }\textbf {\bibinfo {volume} {108}},\ \bibinfo {pages}
  {127204} (\bibinfo {year} {2012})}\BibitemShut {NoStop}%
\bibitem [{\citenamefont {{Beach}}(2004)}]{Beach04a}%
  \BibitemOpen
  \bibfield  {author} {\bibinfo {author} {\bibfnamefont {K.~S.~D.}\
  \bibnamefont {{Beach}}},\ }\href@noop {} {\bibfield  {journal} {\bibinfo
  {journal} {eprint arXiv:cond-mat/0403055}\ } (\bibinfo {year} {2004})},\
  \Eprint {http://arxiv.org/abs/cond-mat/0403055} {cond-mat/0403055}
  \BibitemShut {NoStop}%
\bibitem [{\citenamefont {Yoshitake}\ \emph {et~al.}(2016)\citenamefont
  {Yoshitake}, \citenamefont {Nasu},\ and\ \citenamefont
  {Motome}}]{Yoshioka2016}%
  \BibitemOpen
  \bibfield  {author} {\bibinfo {author} {\bibfnamefont {J.}~\bibnamefont
  {Yoshitake}}, \bibinfo {author} {\bibfnamefont {J.}~\bibnamefont {Nasu}}, \
  and\ \bibinfo {author} {\bibfnamefont {Y.}~\bibnamefont {Motome}},\ }\href
  {\doibase 10.1103/PhysRevLett.117.157203} {\bibfield  {journal} {\bibinfo
  {journal} {Phys. Rev. Lett.}\ }\textbf {\bibinfo {volume} {117}},\ \bibinfo
  {pages} {157203} (\bibinfo {year} {2016})}\BibitemShut {NoStop}%
\bibitem [{\citenamefont {Nasu}\ \emph {et~al.}(2015)\citenamefont {Nasu},
  \citenamefont {Udagawa},\ and\ \citenamefont {Motome}}]{Nasu2015}%
  \BibitemOpen
  \bibfield  {author} {\bibinfo {author} {\bibfnamefont {J.}~\bibnamefont
  {Nasu}}, \bibinfo {author} {\bibfnamefont {M.}~\bibnamefont {Udagawa}}, \
  and\ \bibinfo {author} {\bibfnamefont {Y.}~\bibnamefont {Motome}},\ }\href
  {\doibase 10.1103/PhysRevB.92.115122} {\bibfield  {journal} {\bibinfo
  {journal} {Phys. Rev. B}\ }\textbf {\bibinfo {volume} {92}},\ \bibinfo
  {pages} {115122} (\bibinfo {year} {2015})}\BibitemShut {NoStop}%
\bibitem [{\citenamefont {Modic}\ \emph {et~al.}(2020)\citenamefont {Modic},
  \citenamefont {McDonald}, \citenamefont {Ruff}, \citenamefont {Bachmann},
  \citenamefont {Lai}, \citenamefont {Palmstrom}, \citenamefont {Graf},
  \citenamefont {Chan}, \citenamefont {Balakirev}, \citenamefont {Betts},
  \citenamefont {Boebinger}, \citenamefont {Schmidt}, \citenamefont {Lawler},
  \citenamefont {Sokolov}, \citenamefont {Moll}, \citenamefont {Ramshaw},\ and\
  \citenamefont {Shekhter}}]{Modic20}%
  \BibitemOpen
  \bibfield  {author} {\bibinfo {author} {\bibfnamefont {K.~A.}\ \bibnamefont
  {Modic}}, \bibinfo {author} {\bibfnamefont {R.~D.}\ \bibnamefont {McDonald}},
  \bibinfo {author} {\bibfnamefont {J.~P.~C.}\ \bibnamefont {Ruff}}, \bibinfo
  {author} {\bibfnamefont {M.~D.}\ \bibnamefont {Bachmann}}, \bibinfo {author}
  {\bibfnamefont {Y.}~\bibnamefont {Lai}}, \bibinfo {author} {\bibfnamefont
  {J.~C.}\ \bibnamefont {Palmstrom}}, \bibinfo {author} {\bibfnamefont
  {D.}~\bibnamefont {Graf}}, \bibinfo {author} {\bibfnamefont {M.~K.}\
  \bibnamefont {Chan}}, \bibinfo {author} {\bibfnamefont {F.~F.}\ \bibnamefont
  {Balakirev}}, \bibinfo {author} {\bibfnamefont {J.~B.}\ \bibnamefont
  {Betts}}, \bibinfo {author} {\bibfnamefont {G.~S.}\ \bibnamefont
  {Boebinger}}, \bibinfo {author} {\bibfnamefont {M.}~\bibnamefont {Schmidt}},
  \bibinfo {author} {\bibfnamefont {M.~J.}\ \bibnamefont {Lawler}}, \bibinfo
  {author} {\bibfnamefont {D.~A.}\ \bibnamefont {Sokolov}}, \bibinfo {author}
  {\bibfnamefont {P.~J.~W.}\ \bibnamefont {Moll}}, \bibinfo {author}
  {\bibfnamefont {B.~J.}\ \bibnamefont {Ramshaw}}, \ and\ \bibinfo {author}
  {\bibfnamefont {A.}~\bibnamefont {Shekhter}},\ }\href
  {https://doi.org/10.1038/s41567-020-1028-0} {\bibfield  {journal} {\bibinfo
  {journal} {Nature Physics}\ } (\bibinfo {year} {2020})}\BibitemShut {NoStop}%
\bibitem [{\citenamefont {Kasahara}\ \emph {et~al.}(2018)\citenamefont
  {Kasahara}, \citenamefont {Sugii}, \citenamefont {Ohnishi}, \citenamefont
  {Shimozawa}, \citenamefont {Yamashita}, \citenamefont {Kurita}, \citenamefont
  {Tanaka}, \citenamefont {Nasu}, \citenamefont {Motome}, \citenamefont
  {Shibauchi},\ and\ \citenamefont {Matsuda}}]{Kasahara2018}%
  \BibitemOpen
  \bibfield  {author} {\bibinfo {author} {\bibfnamefont {Y.}~\bibnamefont
  {Kasahara}}, \bibinfo {author} {\bibfnamefont {K.}~\bibnamefont {Sugii}},
  \bibinfo {author} {\bibfnamefont {T.}~\bibnamefont {Ohnishi}}, \bibinfo
  {author} {\bibfnamefont {M.}~\bibnamefont {Shimozawa}}, \bibinfo {author}
  {\bibfnamefont {M.}~\bibnamefont {Yamashita}}, \bibinfo {author}
  {\bibfnamefont {N.}~\bibnamefont {Kurita}}, \bibinfo {author} {\bibfnamefont
  {H.}~\bibnamefont {Tanaka}}, \bibinfo {author} {\bibfnamefont
  {J.}~\bibnamefont {Nasu}}, \bibinfo {author} {\bibfnamefont {Y.}~\bibnamefont
  {Motome}}, \bibinfo {author} {\bibfnamefont {T.}~\bibnamefont {Shibauchi}}, \
  and\ \bibinfo {author} {\bibfnamefont {Y.}~\bibnamefont {Matsuda}},\ }\href
  {\doibase 10.1103/PhysRevLett.120.217205} {\bibfield  {journal} {\bibinfo
  {journal} {Phys. Rev. Lett.}\ }\textbf {\bibinfo {volume} {120}},\ \bibinfo
  {pages} {217205} (\bibinfo {year} {2018})}\BibitemShut {NoStop}%
\end{thebibliography}
%

\clearpage
\section{Supplemental Material}

In this supplemental material section we will first  provide a demonstration of the phase pinning approach  and then show how to implement this idea  for the general Hamiltonian of Eq.~(\ref{Eq:KHM}) of the main text.
Next, we will plot the  uniform susceptibility data presented in the main text on a linear scale so as to emphasis the Curie-Weiss behavior.  
We will then provide further data for the spin-spin correlations in the  zig-zag and stripy phases of the Kitaev-Heisenberg model of Eq.~(\ref{Eq:KHM-2}) of the main text.
Finally we will discuss the  dynamical spin structure factor at higher temperatures  than considered in the main text  for the Kitaev-Heisenberg model of Eq.~(\ref{Eq:KHM-2}) of the main text.

\subsection{The phase pinning approach }
Consider the action: 
\begin{equation}
	   S(\Phi)    =  S_0(\Phi) - \log   \text{Tr} \left[ {\cal T}   e^{-\int_{0}^{\beta}  d \tau   \hat{\ve{c}}^{\dagger} h(\tau)   \hat{\ve{c}}^{\phantom\dagger} } \right] 
\end{equation}
with  
\begin{equation}
\sum_{x,y} \hat{c}^{\dagger}_x  h_{x,y}(\tau)   \hat{c}^{\phantom\dagger}_y  =   \hat{\ve{c}}^{\dagger}  h(\tau)   \hat{\ve{c}}^{\phantom\dagger}.
\end{equation}
Here, $x$, $y$  run over the single particle states. 

We will assume that  we  can find   an  anti-unitary operator:  
\begin{equation}
\hat{T}    =  \hat{K}   \hat{U} 
\end{equation}
 that  commutes with the single body Hamiltonian: 
\begin{equation}
\label{Symm.eq}
	\left[  \hat{\ve{c}}^{\dagger}  h(\tau)   \hat{\ve{c}}^{\phantom\dagger} ,  \hat{T}   \right] = 0 \, \, \,  \forall \, \, \, \tau. 
\end{equation} 
Here  $\hat{U} $  is unitary and   $\hat{K}  $   corresponds to complex conjugation. 

Equation (\ref{Symm.eq}) is equivalent to: 
\begin{equation} 
     \hat{U}^{\dagger} \hat{\ve{c}}^{\dagger}   \hat{U}^{\phantom\dagger}  \overline{h}(\tau)  \hat{U}^{\dagger} \hat{\ve{c}}^{\phantom\dagger} \hat{U}^{\phantom\dagger}   =  \hat{\ve{c}}^{\dagger}  h(\tau)   \hat{\ve{c}}^{\phantom\dagger}
\end{equation}
where $ \overline{h}(\tau) $  denotes the element wise   complex conjugation of the matrix $h(\tau)$. 
Thereby: 
\begin{align}
& \overline{ \text{Tr} \left[ {\cal T}   e^{-\int_{0}^{\beta}  d \tau   \hat{\ve{c}}^{\dagger} h(\tau)   \hat{\ve{c}}^{\phantom\dagger}} \right]  }   =   \text{Tr} \left[ {\cal T}   e^{-\int_{0}^{\beta}  d \tau   \hat{\ve{c}}^{\dagger} \overline{h(\tau)}   \hat{\ve{c}}^{\phantom\dagger}} \right]  =    \\
&  \text{Tr} \left[ {\cal T}   e^{-\int_{0}^{\beta}  d \tau   \hat{U}^{\dagger} \hat{\ve{c}}^{\dagger}  \hat{U}^{\phantom\dagger} \overline{h(\tau)}   
 \hat{U}^{\dagger}  \hat{\ve{c}}^{\phantom\dagger}\hat{U}^{\phantom\dagger}}   \right]  = 
 \text{Tr} \left[ {\cal T}   e^{-\int_{0}^{\beta}  d \tau   \hat{\ve{c}}^{\dagger} h(\tau)   \hat{\ve{c}}^{\phantom\dagger}} \right].   \nonumber 
\end{align}
In the above,  $ \hat{U}^{\dagger}  \hat{\ve{c}}^{\phantom\dagger}\hat{U}^{\phantom\dagger} $  corresponds to a canonical transformation of the $  \hat{\ve{c}}^{\phantom\dagger} $ fermion operator  such that the  trace remains invariant. 
Hence,  $\text{Tr} \left[ {\cal T}   e^{-\int_{0}^{\beta}  d \tau   \hat{\ve{c}}^{\dagger} h(\tau)   \hat{\ve{c}}^{\phantom\dagger}} \right]$  is real  and
\begin{equation}
	\text{Im}S = 0, \pi. 
\end{equation}

\subsection{Optimal AFQMC formulation of the generalized Kitaev model}
Here we show that   we can apply the phase pinning method to the generalized Kitaev model  of Eq.~(\ref{Eq:KHM}) of the main text.
We start by adopting  a fermion representation of the spin-1/2 degree of freedom: $ \hat {\ve{S}}_{i} = \frac{1}{2} \hat{\ve{ f}}^{\dagger}_{i}   \hat{ \ve{\sigma}}  \hat{ \ve{f}} ^{\phantom\dagger}_{i}  $ where $\hat{\ve{f}}^{\dagger}_{i}  \equiv  \left(\hat {f}^{\dagger}_{i,\uparrow}, \hat f^{\dagger}_{i,\downarrow} \right) $ is a two-component fermion with constraint $\hat{\ve{f}}^{\dagger}_{i}  \hat{\ve{f}}^{\phantom\dagger}_{i}   = 1$.  
Let us now  relax the constraint  on the Hilbert space, and enforce by adding  a Hubbard $U$ term on each site.  The Hamiltonian that we will simulate reads: 
\begin{eqnarray}
\label{Eq:HQMC-sm}
\hat{H}_{{\rm QMC}}    &= & \sum_{ i ,j, \alpha, \beta }\frac{|\Gamma_{i,j}^{\alpha,\beta} |}{2}  \left( \hat{S}_{i}^{\alpha}  + \frac{\Gamma_{i,j}^{\alpha,\beta}} {|\Gamma_{i,j}^{\alpha,\beta} |}   \hat{S}_{j}^{\beta}  \right)^2   \nonumber \\
& & - \sum_{ i ,j}  \frac{J_{i,j}}{8} \left(  \left(  \hat{D}^{\dagger}_{i,j}  + \hat{D}^{\phantom \dagger}_{i,j}  \right)^2+ \left(i \hat{D}^{\dagger}_{i,j}   -i \hat{D}^{\phantom\dagger}_{i,j}  \right)^2  \right) \nonumber \\
& &+U  \sum_{i} \left( \hat{\ve{f}}^{\dagger}_i \hat{\ve{f}}^{\phantom\dagger}_i  - 1 \right)^2,
\end{eqnarray}
where $ \hat{D}^{\dagger}_{i,j} = \hat{\ve{f}}^{\dagger}_i \hat{\ve{f}}^{\phantom\dagger}_{j} $.     
 It is important to note that 	$\left[  \left(  \hat{\ve{f}}^{\dagger}_{i}  \hat{\ve{f}}^{\phantom\dagger}_{i}  -1  \right)^2 ,  \hat H_{\rm{QMC}}   \right]  = 0 $ such that the  $\hat{\ve{f}}$-fermion parity  $ (-1)^{\hat{\ve{f}}^{\dagger}_{i} \hat{\ve{f}}^{\phantom\dagger}_{i} }  $  is a local conserved quantity  and that the constraint is very efficiently imposed. We will discuss this point at the end of the section.    In the odd parity sector favored by the  repulsive  Hubbard interaction, $ \left. \hat{H}_{\rm{QMC}} \right|_{(-1)^{ \hat{\ve{f}}^{\dagger}_{i}  \hat{\ve{f}}^{\phantom\dagger}_{i} }   = -1 }= \hat{H} + C $ where $C$ is a constant.

 The above form  in terms of perfect squares can be implemented in the ALF-implementation of the  auxiliary field QMC (AFQMC)  algorithm.       As  mentioned in the main text, the $J_{i,j}$  exchange constants are non-frustrating. This means that    we can find a set of Ising spins, $s_i= \pm 1$,  such that for  each bond with $J_{i,j}  \neq 0 $,  $J_{i,j} s_i s_j < 0$.  Hence, 
 \begin{equation}
 	  J_{i,j}  = |  J_{i,j} | \left(- s_i s_j \right).  
 \end{equation}
 After  Trotter decomposition and Hubbard-Stratonovich transformation the grand canonical  partition function reads: 
 \begin{eqnarray}
    Z = & & {\rm{Tr}} \left[ e^{- \beta \hat{H}_{\rm{QMC}}}\right]    \propto     \nonumber \\
      & & \int   D \left\{ \chi_{i,j}^{\alpha,\beta}(\tau) , {\rm{Re}}Z_{i,j}(\tau),  {\rm{Im}}Z_{i,j}(\tau) , \lambda_i(\tau) \right\}      \nonumber \\
      & & \times e^{- S\left( \left\{  \chi_{i,j}^{\alpha,\beta}(\tau), Z_{i,j}(\tau), \lambda_i(\tau)   \right\} \right)}.
 \end{eqnarray}
 For given field configuration,  $\chi_{i,j}^{\alpha,\beta}(\tau),  \lambda_i(\tau)  \in \mathbb{R} $  and $Z_{i,j}(\tau)  \in \mathbb{C} $,  the action is given by: 
 \begin{eqnarray}	
S \left( \left\{ \chi,Z,\lambda \right\}  \right) =   && 
     \int_{0}^{\beta} d \tau  \left[  \sum_{i ,j, \alpha,\beta}  \frac{ \left( \chi_{i,j}^{\alpha,\beta}(\tau)\right)^2 }{|\Gamma_{i,j}^{\alpha,\beta} |}     \right.  \nonumber \\
    &&  \left. + \sum_{i,j} \frac{ |   Z_{i,j}(\tau)|^2 } { 4 |J_{i,j} | }  + \sum_{i} \frac{ \lambda_i(\tau)^2 }{ 2U } \right]   \nonumber  \\  & &    -  \ln {\rm{Tr}}  {\cal T} e^{-\int_{0}^{\beta } d \tau \hat{H}(   \left\{ \chi,Z,\lambda \right\})}
\end{eqnarray}
with 
\begin{eqnarray}
	\hat{H} (   \left\{ \chi,Z,\lambda \right\} )    &= & \sum_{ i,j, \alpha, \beta  }  i  \chi_{i,j}^{\alpha,\beta}(\tau) \left( \hat{S}_{i}^{\alpha}  + \frac{\Gamma_{i,j}^{\alpha,\beta}} {|\Gamma_{i,j}^{\alpha,\beta} |}   \hat{S}_{j}^{\beta}  \right)   \nonumber \\
& &  + \sum_{ i,j \delta } \sqrt{-s_i s_j} \left(   Z_{i,j} (\tau)  \hat{D}^{\dagger}_{i,j}  + \overline{Z_{i,j} (\tau) } \hat{D}^{\phantom \dagger}_{i,j}  \right)  \nonumber \\
& &+ \sum_{i} i \lambda_i(\tau) \left( \hat{\ve{f}}^{\dagger}_i \hat{\ve{f}}^{\phantom\dagger}_i  - 1 \right). 
\end{eqnarray}
In the above, it is understood that the first  sum runs over   bonds  and spin indices  where   $ \Gamma_{i,j}^{\alpha,\beta} \neq 0$.  Similarly the second sum runs over bonds where  $J_{i,j}$  does not vanish. 
 Now  consider the anti-unitary transformation  
\begin{equation}
\label{Symm.eq2}
	  \hat{T} \alpha \hat{f}^{\dagger}_{\ve{i},\sigma} \hat{T}^{-1} =  \overline{\alpha}  s_i \hat{f}^{\phantom\dagger}_{\ve{i},\sigma}
\end{equation}
where $\alpha $ is a complex number.  
One will show that 
\begin{equation}	
	 \hat{T}  \hat{H} (   \left\{ \chi,Z,\lambda \right\} )   \hat{T}^{-1}  =   \hat{H} (   \left\{ \chi,Z,\lambda \right\} )   
\end{equation}
such that for this formulation 
\begin{equation}
\label{Eq:Quant}
	   {\rm{Im}} S  \left( \left\{ \chi,Z,\lambda \right\}  \right)     =  0, \pi.
\end{equation}

 \begin{figure}
\centering
\centerline{\includegraphics[width=0.48\textwidth]{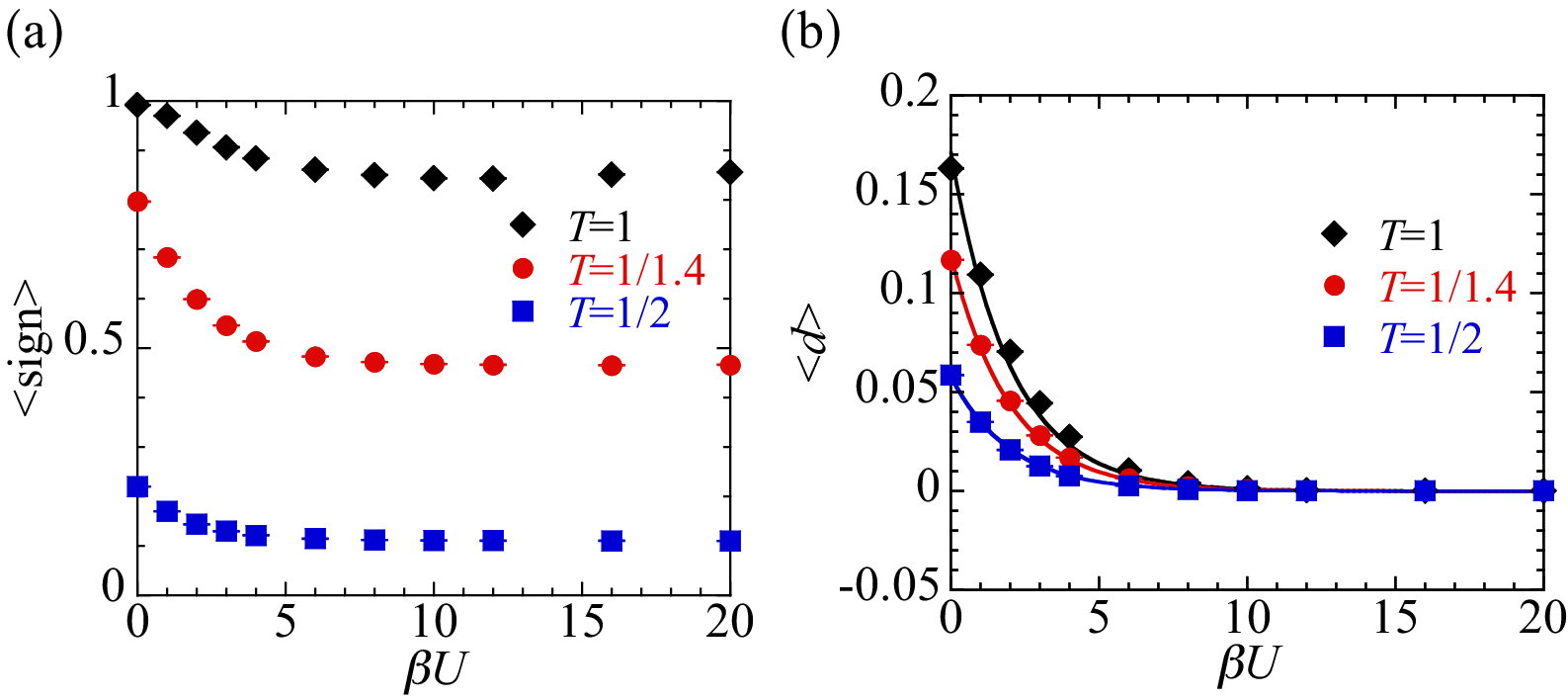}}
\caption{ $\beta U$ dependence of (a) average sign $\langle \text{sign} \rangle$ and (b) double occupancy
 $d=\langle  \hat {f}^{\dagger}_{i,\uparrow} \hat {f}^{}_{i,\uparrow} \hat {f}^{\dagger}_{i,\downarrow} \hat {f}^{}_{i,\downarrow} \rangle $ for different temperatures $T=1/\beta$. Here, $\varphi/\pi=0.2$ (see text) and lattice size $V=32$.
The solid lines in (b) are the results of the fitting Eq.(\ref{eq:d-scaling}).
}
\label{fig:Fig9} 
\end{figure}

\begin{figure}
\centering
\centerline{\includegraphics[width=0.45\textwidth]{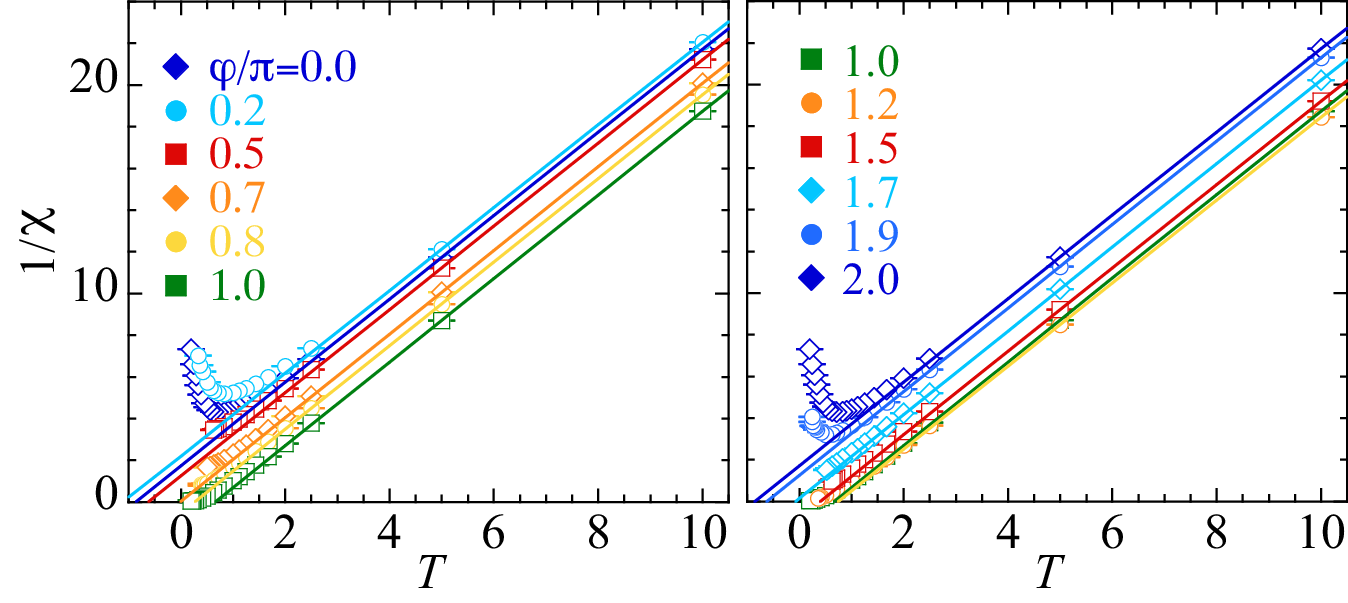}}
\caption{
$T$ dependence of inverse uniform spin susceptibilities $1/\chi$ at different values of $\varphi/\pi$.
The solid lines are a fit   to Curie-Weiss law  in the range  $T  \in [5, 10]$.
}
\label{fig:unisus-linearplot} 
\end{figure}

We note that when the generalized Kitaev term is set to zero,  the action for each field configuration, and the model, have an additional SU(2) spin symmetry.  This implies  that the fermion determinant factorizes in up and down spin sectors.   Owing to the SU(2) spin symmetry the fermion determinants are identical in each spin sector. The anti-unitary  transformation of Eq.~(\ref{Symm.eq2})  can be applied in each spin sector to show that the fermion determinant is real.    Hence in this case, there is no  sign problem  since the weight is given by the square of a real number. 

We conclude this  section by   discussing the  convergence to the physical Hilbert space. Since, as mentioned above,  $\left[  \left(  \hat{\ve{f}}^{\dagger}_{i}  \hat{\ve{f}}^{\phantom\dagger}_{i}  -1  \right)^2 ,  \hat H_{\rm{QMC}}   \right]  = 0 $     one can  show that 
\begin{equation}
\label{eq:d-scaling}
	\left< \left(  \hat{\ve{f}}^{\dagger}_{i}  \hat{\ve{f}}^{\phantom\dagger}_{i}  -1  \right)^2     \right>    \propto e^{-\beta U/2}.
\end{equation}
Owing to   the invariance of the action  under the particle-hole symmetry of Eq.~(\ref{Symm.eq2}), $  \left<  \hat{\ve{f}}^{\dagger}_{i}  \hat{\ve{f}}^{\phantom\dagger}_{i}   \right> = 1 $  such  that 
\begin{equation}
   \left< \left(  \hat{\ve{f}}^{\dagger}_{i}  \hat{\ve{f}}^{\phantom\dagger}_{i}  -1  \right)^2     \right>      =  2 \langle  \hat {f}^{\dagger}_{i,\uparrow} \hat {f}^{}_{i,\uparrow} \hat {f}^{\dagger}_{i,\downarrow} \hat {f}^{}_{i,\downarrow} \rangle. 
\end{equation}
The double occupancy is plotted in Fig.~\ref{fig:Fig9} (b)    and as apparent follows the predicted exponential form.     Clearly values of $\beta U = 10$  suffice to guarantee convergence to the physical  Hilbert space.    It is vey interesting to consider the average sign as a function of $\beta U$.  Generically,  the  sign  decays exponentially with  inverse temperature.    In contrast to this general expectation, Fig.~\ref{fig:Fig9} (a), shows that the  average sign converges to a constant.

\begin{figure}
\centerline{\includegraphics[width=0.42\textwidth]{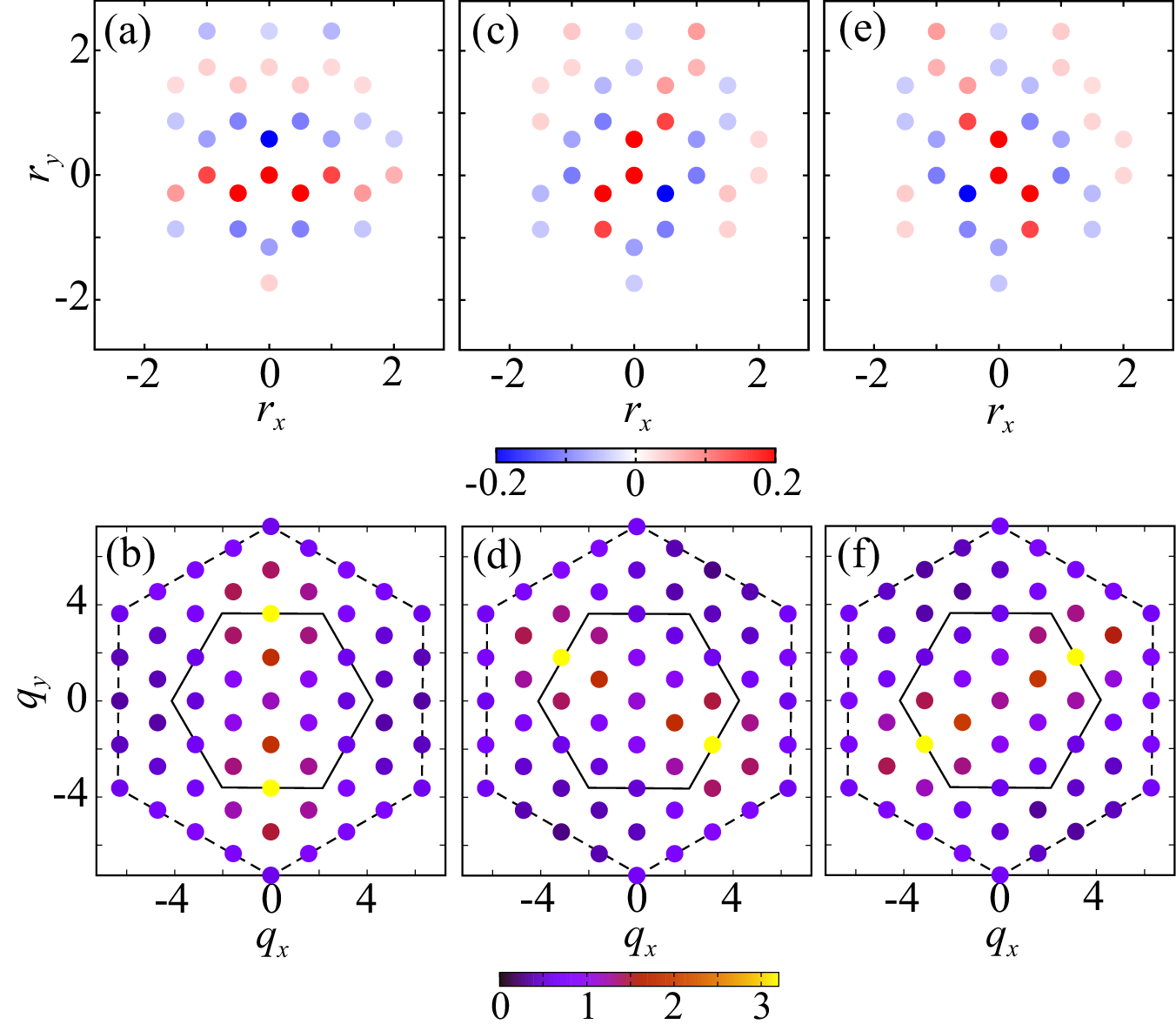}}
\caption{\label{fig:ZZ-SSsusq}   
Real-space spin-spin correlations $\langle  \hat{S}^{\alpha}_{\vec{r}} \hat{S}^{\alpha}_{\vec{0}} \rangle$ (top panel) and momentum  resolved spin susceptibility
$\chi_{\alpha}(\vec{q})$ (bottom panel) in the first (solid) and second (dashed line) Brillouin zones.   Here we consider  $\varphi/\pi=0.8$ and $T=1/2.6$.
(a)-(b) $\alpha=1$, (c)-(d) $\alpha=2$, and (e)-(f) $\alpha=3$.
}
\end{figure}
\begin{figure}
\centerline{\includegraphics[width=0.42\textwidth]{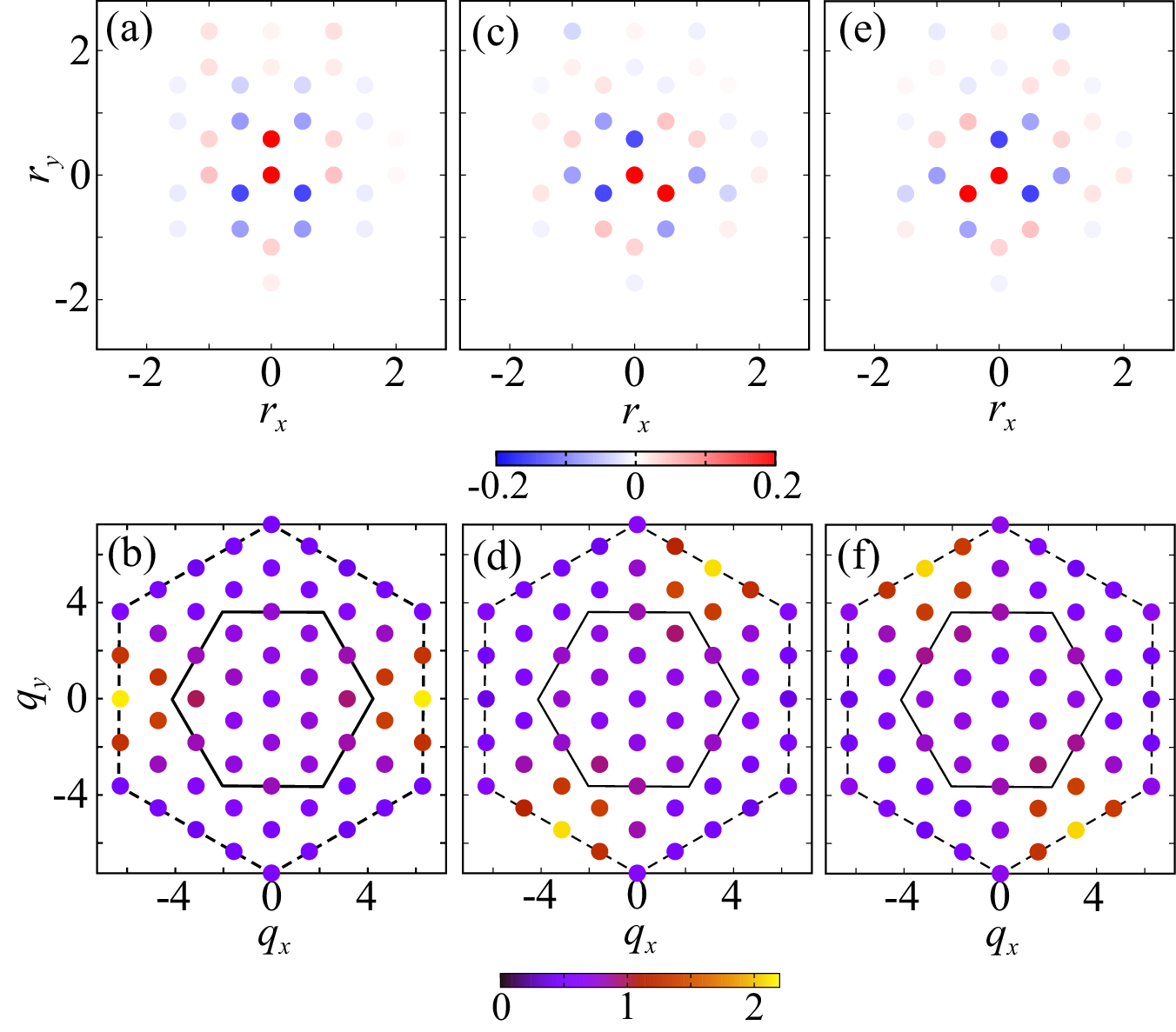}}
\caption{\label{fig:SP-SSsusq}
Real-space spin-spin correlations $\langle  \hat{S}^{\alpha}_{\vec{r}} \hat{S}^{\alpha}_{\vec{0}} \rangle$ (top panel) and momentum  resolved spin susceptibility
$\chi_{\alpha}(\vec{q})$ (bottom panel) in the  first (solid) and second (dashed line) Brillouin zones.   Here we consider  $\varphi/\pi=1.7$ and $T=1/1.9$.
(a)-(b) $\alpha=1$, (c)-(d) $\alpha=2$, and (e)-(f) $\alpha=3$.
}
\end{figure}

\begin{figure}
\centering
\centerline{\includegraphics[width=0.45\textwidth]{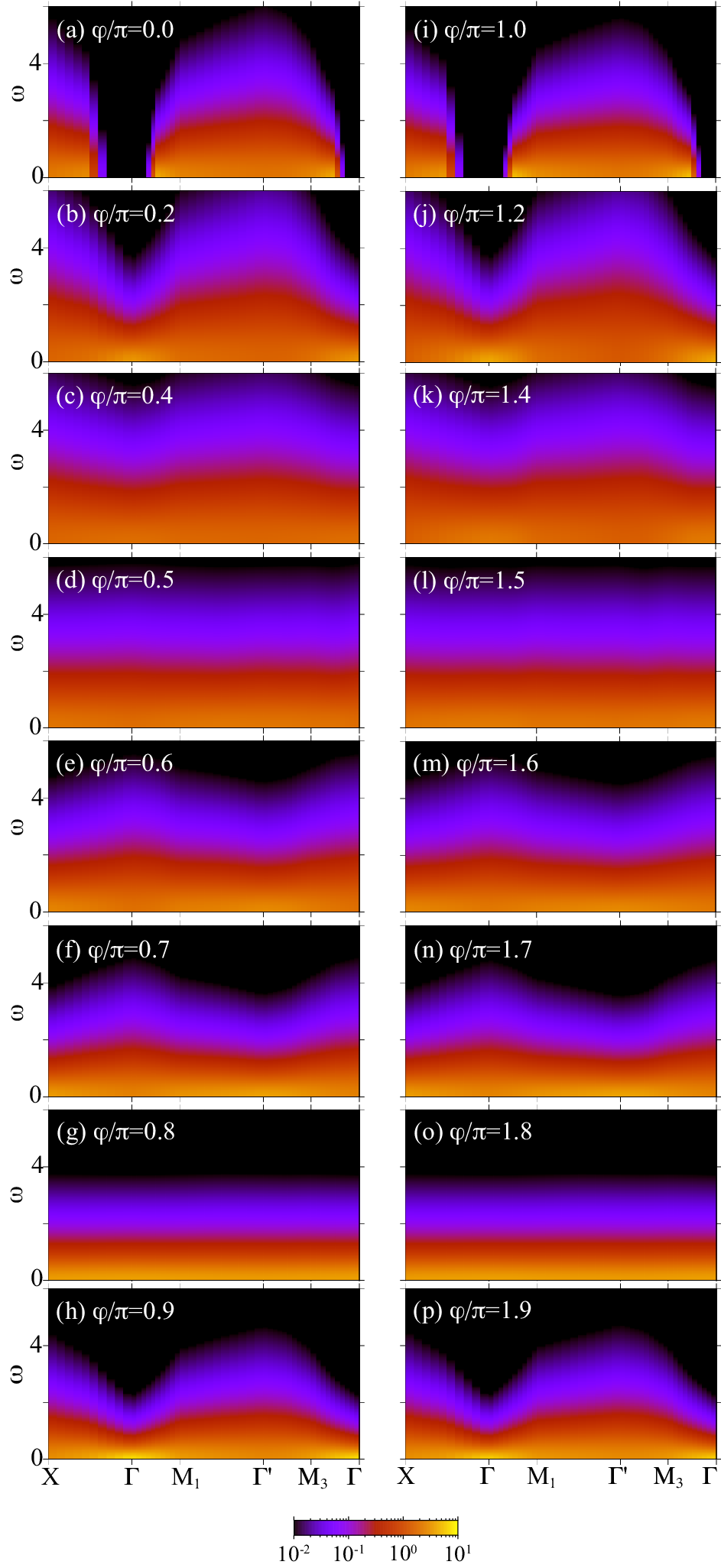}}
\caption{\label{fig:sus0.1}  Dynamical spin structure factor $C(\boldsymbol{q}, \omega)$ at different values
of $\varphi/\pi$. Here, $T=10$. Results used here correspond to scans along the red 
 line of Fig.~\ref{fig:Fig1}(b) of the main text.
}
\label{fig:Cqomega3}  
\end{figure}

\subsection{Curie-Weiss behaviors}
In the main text our QMC results for the uniform spin susceptibilities $\chi$ as a function of the angle $\varphi$ and temperature support the departure from the high-temperature Curie law in the ordered and disordered phases inherent to the Kitaev-Heisenberg model of Eq.~(\ref{Eq:KHM-2}) of the main text.
At very high temperatures  local correlations are impaired so that the Curie law is obeyed.  With decreasing temperatures local correlations develop   and one can expect the Curie law to give way to  Curie-Weiss one, at least in an intermediate temperature range. 
In Fig.~\ref{fig:unisus-linearplot}, we plot our inverse susceptibility data, $1/\chi$,  presented in the main text on a linear scale. 
For all values of $\varphi$, $1/\chi$  indeed follows the predicted Curie-Weiss form $1/\chi=(T-\Theta_{\rm cw})/C$   in an intermediate temperature range.  
Note that as a function of $\varphi$ the sign of the Curie-Weiss temperature  $\Theta_{\rm cw}$  changes  from negative to positive reflecting the sign of the dominant local exchange coupling.

\subsection{Spin correlations in the zig-zag and stripy phases}
The Kitaev-Heisenberg model of Eq.~(\ref{Eq:KHM-2}) of the main text remains invariant under  combined $2  \pi/3$  rotations  and  a permutation of the elements of the spin vector $ \left( \hat{S}^{1}_{\vec{r}}, \hat{S}^{2}_{\vec{r}},\hat{S}^{3}_{\vec{r}} \right) $.   As  a consequence  the ordering pattern in the zig-zag (Fig.~\ref{fig:ZZ-SSsusq}) and stripy (Fig.~\ref{fig:SP-SSsusq}) phases,  rotate by a $2\pi/3$  angle when measuring for instance   $\langle \hat{S}^{2}_{\vec{r}}  \hat{S}^{2}_{\vec{0}}  \rangle $   instead of  $\langle \hat{S}^{1}_{\vec{r}}  \hat{S}^{1}_{\vec{0}}  \rangle $.   Figures~\ref{fig:ZZ-SSsusq}  and \ref{fig:SP-SSsusq}   confirm this, thus providing a benchmark for our code.

\subsection{High-temperature spin dynamics}
In order to capture finite temperature properties of ordered and spin-liquid ground states inherent to the Kitaev-Heisenberg model of Eq.~(\ref{Eq:KHM-2}) of the main text, we computed the dynamical spin structure factor at  different temperatures.
In our QMC simulations, the dynamical spin structure factor $C(\boldsymbol{q}, \omega)$  of  Eq.~(\ref{CSq}) of the main text is obtained via the analytic continuation of the imaginary-time-displaced spin correlation functions.   We used the Algorithms for Lattice Fermions (ALF) \cite{ALF_v1}  implementation of the stochastic  analytical  continuation \cite{Beach04a}. 

In  the main text  we show that  $C(\boldsymbol{q}, \omega)$ at $T=1/1.6$  and as a function of $\varphi$ picks up the distinct  finite temperature  features of the   ordered and disordered phases of the Kitaev-Heisenberg model. 
As the temperature increases,  local correlations are impaired so that  $C(\boldsymbol{q}, \omega)$ is expected to become  ${\boldsymbol{q}}$-independent with  spectral weight centered around low frequencies. 
Figure~\ref{fig:Cqomega3} shows results at higher temperatures, $T=10$.     Consider $\varphi/\pi = 0.8$  corresponding to the zig-zag phase.  Comparison of the high temperature data in Fig.~\ref{fig:Cqomega3}  with that of the lower temperature data in Fig.~\ref{fig:Cqomega2}  of the main text shows spectral weight shifting for low to high energies and the emergence of distinct $\vec{q}$  dependence.  Note that the angles $\varphi/\pi = 0, 1$ stand apart  due to the enhanced SU(2) spin symmetry.   For these angles the total spin is a conserved quantity  such that the  dynamical spin structure factor at the  $\ve{\Gamma}$ point and at any  temperature  is given by a Dirac $\delta$-function in frequency.

\end{document}